\documentclass[aps,prd,footnote]{revtex4-2}
\usepackage{blindtext}
\usepackage{graphicx}
\usepackage{amssymb}
\usepackage{lipsum}
\usepackage{physics}
\usepackage{mathrsfs}
\usepackage{MnSymbol}
\usepackage{soul}
\usepackage{xcolor}
\DeclareMathOperator{\arcsinh}{arcsinh}

\newcommand{\be}{\begin{equation}}
\newcommand{\ee}{\end{equation}}

\newcommand{\PR}[1]{\ensuremath{\left[#1\right]}}
\newcommand{\PC}[1]{\ensuremath{\left(#1\right)}}
\newcommand{\chav}[1]{\ensuremath{\left\{#1\right\}}}
\newcommand{\bes}{\begin{subequations}}
\newcommand{\ees}{\end{subequations}}

\begin{document}
\title{Geometrically Constrained Localized Configurations: First-Order Framework\\ and Analytical Solutions}

\author{D. Bazeia$^1$
\!\!\footnote{Corresponding author. Email: dbazeia@gmail.com}}
\author{{M. A. Feitosa$^{1,2}$}}
\author{R. Menezes$^3$}
\author{G. S. Santiago$^1$}

\affiliation{${^1}\!\!$ {Departamento de Física, Universidade Federal da Paraíba, João Pessoa, Paraíba, Brazil}}
\affiliation{${^2}\!\!$ {Instituto de Física, Universidade Federal de Goiás, Gioânia, Goiás, Brazil}}
\affiliation{${^3}\!\!$ {Departamento de Ciências Exatas, Universidade Federal da Paraíba, Rio Tinto, Paraíba, Brazil}}

\begin{abstract}
\bigskip

\bigskip

This work deals with the presence of topological structures in models of two real scalar fields in the two-dimensional spacetime. The subject concerns the presence of a geometric constriction, which appears with a modification of the kinetic term of one of the two fields. We elaborate on the construction of a first-order framework, which directly contributes to find analytical solutions. We describe several distinct possibilities, in particular, the case where the first-order equations do not separate. This is much harder, but we use the integrating factor to deal with analytical configurations. The proposed methodology help us deal with localized structures of both the Néel and Bloch type very naturally, and we end the work suggesting some possibilities of applications in distinct areas of nonlinear science.
\end{abstract}

\maketitle

\section{Introduction}

The study of localized structures known as solitons and soliton-like configurations in homogeneous and inhomogeneous systems is of great interest in nonlinear science in general; see, e.g. Ref. \cite{KM} and references therein. In high energy physics, in particular, the search for localized structures constructed under the presence of scalar and gauge fields is a very active area of research. The books \cite{Vi,Ma,Va,Sc} deal with the subject and describe interesting distinct possibilities of obtaining and using localized structures to investigate specific issues of practical importance in field theory.

We have learned from the studies developed mainly in the last 60 years, that real scalar fields are  important to describe localized structures in one spatial dimension. In this case, they are called kinks, but we can also have vortices and magnetic monopoles, vortices being two-dimensional solutions that appear under the presence of a complex scalar field minimally coupled to an Abelian gauge field \cite{NO}. They may appear in several distinct situations, for instance, in fluids \cite{F}, in superconductors \cite{Su}, and in Bose-Einstein condensates \cite{BEA,BEB}. Furthermore, magnetic monopoles can be found in three spatial dimensions in the presence of non Abelian gauge symmetries; they were first considered in Refs. \cite{tH,Po}; see also Refs. \cite{Sch,Mil}. They may also appear in condensed matter, as emergent particles in spin ice systems \cite{SI,SI2}.  In the case of kinks, they may sometimes behave as solitons, as happens with solutions of the sine-Gordon model, which preserve their shape under propagation, and recover them after collisions with other such localized configurations (see, e.g., \cite{Malo} for more information on the subject), but they may also appear in non-integrable models like the $\phi^4$ model in the presence of spontaneous symmetry breaking, which engenders the well-known kinklike form, analytically represented by the hyperbolic tangent function. Here, however, they may present very interesting output possibilities after colliding with one another \cite{Camp}.

Kinks, vortices and magnetic monopoles are important localized structures, and they all attain topological features.  
In the present work, however, we shall deal with kinks, which are spatially localized one-dimensional configurations that have simple topological properties. Although kinks are perhaps the simplest of the known localized structures, we shall bring novelties considering models described by two real scalar fields, concentrating mainly on the  possibility firstly considered in \cite{Const} to constrain the geometry of the system. In this case, in the two-field model, one takes one of the fields to work independently and to further act to constrain the geometry of the other field, contributing to modify the center \cite{Const,sala} or the tail \cite{tail} of the corresponding kinklike solution. An important motivation associated to this description is that the procedure allows for the presence of analytical results that engender the possibility of adding a plateau at the center of the configuration, in this sense  connecting the exact solutions to the experimental observation of a similar behavior in magnetic material in the presence of constrained geometry \cite{prbrc}.  his is an important fact, linking magnetic configurations in constrained geometries to kinks in relativistic models in the presence of kinematic modification.

Kinks in models described by scalar fields were studied before in Refs.  \cite{DHN,Montonen,STB,Raj,Ruck} with distinct motivation, in particular, searching for the presence of  analytical solutions. For instance, in \cite{Raj} the author developed a method to solve coupled scalar fields theories, and in \cite{Ruck} some analytical solutions were constructed for models engendering the discrete $Z(2)$ and $Z(3)$ symmetries. Kinks have also been considered in several other situations, for instance, to help us understand the existence of metallic polymers like polyacetylene \cite{JR,SSH,Hee}, and the presence of topological twistons in polymeric chains such as  polyethylene \cite{Mansfield,Ventura}. As shown in \cite{JR} and also in \cite{GW}, fermions in the presence of localized structures may engender the interesting property that allows for the presence of 
fractional quantum numbers. This behavior has been further explored in \cite{AMo}, when the kinklike background is geometrically deformed, giving rise to several bound states in the fermionic gap,  an effect that is different from the standard case discussed in \cite{JR}, which showed the presence of a single mode, just the zero mode inside the fermionic gap. Moreover, the effects of an electric current on the polarity of a domain wall configuration with the kinklike profile was also investigated experimentally in Ref. \cite{All}. Here one recalls that in the theoretical study developed in \cite{AMo}, the authors also suggested the possibility to further control the wall polarity  though the presence of geometric constrictions in the magnetic material. Localized structures with the kinklike profile are also of direct importance to describe networks of topological configurations \cite{Net1,Net2,Net3}, and to tile the plane with \cite{G,Sa} or without \cite{BB} supersymmetry. Models described by scalar fields that support kinklike configurations can also be used in collisions, as considered, for instance, in the scattering of kinks in the standard $\phi^4$ model \cite{Camp}, and more recently in the $\phi^6$ model \cite{P6}, in the case where the kink engenders long range tail \cite{LR}, in the $\phi^4$ model in the presence of impurity \cite{IM}, and in the case of spectral walls in multifield kink dynamics \cite{SW}. They can also be considered in collision in the case of a model that support the presence of geometric constrictions \cite{GC}.

The concrete possibility of modifying the center or the tail of the kinklike configuration \cite{Const,sala,tail}, and its direct connection with the geometric constriction found in Ref. \cite{prbrc}, have motivated us to further explore the subject, adding more results and enlarging the scope of the scalar field models,  bringing interesting new results to this specific subject. Another motivation is related to the fact that one dimensional kinklike solutions can be immersed into two or three spatial dimensions to form domain ribbons or domain walls, respectively, which are also used to describe planar interfaces or spatial domains of current interest in nonlinear science. Domain walls, for instance, may be of the Néel or Bloch type, with the wall having nontrivial internal structure which may help us understand the presence of magnetic domain in magnetic materials \cite{MD}. They can also be used to model the magnetization in magnetic materials that support configurations known as skyrmions; see, e.g., Ref. \cite{sky} and references therein. This possibility can be implemented with scalar fields, following the lines of Refs. \cite{sky1,sky2}. 
An important issue related to the description of the magnetization in magnetic materials in the condensed matter framework is that it usually includes the Dzyaloshinkii-Moriya (DM) interaction  \cite{D,M}; see also Refs. \cite{MD,DMInew} for more details on this issue. In particular, in the recent review \cite{DMInew} one finds several important consequences of the DM interaction in magnetic elements. There we see that the DM interaction appears as an antisymmetric exchange interaction that arises due to spin orbit coupling in a crystal structure that breaks inversion symmetry, thus favoring the formation of chiral spin structures. In fact, compared to the Heisenberg exchange interaction $\vec{S}_i\cdot\vec{S}_j$ between dimensionless spins $\vec{S}_i$ and $\vec{S}_j$, which may favor the parallel alignment of spins, the DM contribution is guided by ${\vec{S}_i\times \vec{S}_j}$, which is clearly antisymmmetric and may feed the presence of chiral structure. One should notice, however, that the DM interaction requires the presence of spin-orbit coupling in a lattice structure that breaks inversion symmetry \cite{DMInew}. Although this is out of the scope of the present work, we think the several models investigated analytically in Secs. \ref{R} and \ref{IV} will motivate new research on the subject, in particular, considering different modifications in the dynamics of the two fields, to make them simulate the DM interaction. In the same direction, another topic of current interest is that the geometric modification introduced in the models investigated below may be generalized to control the motion of Bloch and Néel walls and skyrmions in magnetic materials. As one knows, the motion of the localized structure may be caused by the presence of currents, external fields and/or curvature; see, e.g. Refs. \cite{PRL,Toma,PRB} and references therein for more information on this subject. In the case of a curved spacetime, in particular, it is very natural to include curvature in the relativistic models described in the present work, so we shall further comment on these possibilities in Sec. \ref{E}.

To deal with the above motivations, we organize the work as follows. In the next Sec. \ref{M} we focus on the methodology, that is, we introduce the two-field model and describe the general way to obtain first-order differential equations that solve the equations of motion. The next step is to investigate the  models to be introduced in Sec. \ref{R}, finding analytical solutions and describing some of their specific features. We work with several possibilities, with one of the two first-order equations being independent of the other field. In Sec. \ref{IV} we consider another possibility, where the first-order equations are not separable anymore, leading to the perhaps hardest situation. In this case, we search for the integrating factor to find analytical solutions. We close the work in Sec. \ref{E}, where we summarize the main results and comment on the possibility of investigating other related issues of current interest. 

\section{Methodology}
\label{M}

Let us deal with systems described by two real scalar fields. We develop the methodology to be used in this work by firstly reviewing the possibility of describing two-field models with standard dynamics, in the presence of first-order differential equations that solve the equations of motion. We then move on and investigate the novel possibility, changing the dynamics of the model, to get to the case considered in Ref. \cite{Const}, which also support first-order equations. 

\subsection{Standard Dynamics}

We study $1+1$ dimensional systems with the diagonal part of the metric given by $(+,-)$. Also, we consider the presence of the two real scalar fields, $\phi=\phi(x,t)$ and $\chi=\chi(x,t)$. In the case of standard dynamics, we introduce the Lagrangian density
\be
    \mathcal{L} = \frac{1}{2}\partial_{\mu}\phi\partial^{\mu}\phi + \frac{1}{2}\partial_{\mu}\chi\partial^{\mu}\chi - V(\phi, \chi),
\ee
where $\mu=0,1$ is used to describe the Minkowski spacetime, with $x^\mu=(x^0=t, x^1=x)$, such that  $\partial_0\phi=\partial\phi/\partial t=\dot\phi$ and  $\partial_1\phi=\partial\phi/\partial x=\phi^\prime$. The potential is $V=V(\phi,\chi)$, and may contain self-interactions and interactions between the two field.  The equations of motion are given by 
\begin{align}
\begin{split}
 \partial_{\mu}\partial^{\mu}\phi + \frac{\partial V}{\partial\phi} = 0,\;\;\;\;\;\;\;\;
\partial_{\mu}\partial^{\mu}\chi + \frac{\partial V}{\partial\chi} = 0.
\end{split}
\end{align}
As usual, we shall consider the case of static configurations. Thus, the above equations become
\begin{align}
\begin{split}
\label{Second Order A}
    \phi '' = V_{\phi},
    \;\;\;\;\;\;\;\;\;
    \chi '' = V_{\chi},
\end{split}
\end{align}
where $V_{\phi} = \partial V /\partial \phi$ and $V_{\chi} = \partial V /\partial \chi$. In order to establish a first-order framework using the Bogomol'nyi procedure, as discussed in \cite{Bogomol'nyi}, we include an auxiliary function $W(\phi, \chi)$ such that the potential is defined as
\begin{equation}
\label{Potential}
    V(\phi, \chi) = \frac{1}{2}W_{\phi}^2 + \frac{1}{2}W_{\chi}^2,
\end{equation}
where $W_{\phi} = \partial W/\partial \phi$ and $W_{\chi} = \partial W/\partial \chi$. As a consequence, the equations of motion \eqref{Second Order A} now become
\begin{align}
\begin{split}
 \frac{d^2\phi}{dx^2} = W_{\phi}W_{\phi\phi} + W_{\chi}W_{\chi\phi} ,
\;\;\;\;\;\;\;\;
\frac{d^2\chi}{dx^2} = W_{\phi}W_{\phi\chi} + W_{\chi}W_{\chi\chi},
\end{split}
\end{align}
and the energy density can be written in the form
\begin{equation}
    \rho = \frac12\PC{\frac{d\phi}{dx}}^2 +\frac12\PC{\frac{d\chi}{dx}}^2 + \frac12 W_{\phi}^2  +\frac12W_{\chi}^2=\frac12\PC{\frac{d\phi}{dx} \mp W_{\phi}}^2 + \frac12\PC{\frac{d\chi}{dx} \mp W_{\chi}}^2 \pm \frac{dW}{dx}.
\end{equation}

If the fields obey the first-order equations
\begin{align}
\begin{split}
\label{PrimeiraOrdem}
 \frac{d\phi}{dx} = \pm W_{\phi},
\;\;\;\;\;\;\;\;
\frac{d\chi}{dx} = \pm W_{\chi},
\end{split}
\end{align}
the energy gets to its lower bound $E_{B} = \vert W(\phi (\infty), \chi (\infty)) - W(\phi (-\infty), \chi (-\infty)) \vert$. The first-order equations solve the equations of motion when one imposes that $W_{\phi\chi}=W_{\chi\phi}$. This is the Bogomol'nyi procedure, and it implies that the minimum energy value only depends on the asymptotic behavior of $W(\phi,\chi)$, for the solutions that obey the first-order equations.

{\bf {A.1. The BNRT model --}} To illustrate the above procedure, let us first deal with the model considered in Ref. \cite{BSR,BNRT}. This means that one should consider
\begin{equation}
\label{BRNT_W}
    W(\phi, \chi) = \phi - \frac{1}{3}\phi^3 - r\phi\chi^2,
\end{equation}
where $r$ is a real parameter which controls the interaction of the fields. Using \eqref{PrimeiraOrdem} with the plus signals, the first-order equations become
\begin{align}
\label{EquationofmotionBNRT}
\begin{split}
 \frac{d\phi}{dx} = 1 - \phi^2 -r\chi^2,
\;\;\;\;\;\;\;\;
\frac{d\chi}{dx} = -2r\phi\chi,
\end{split}
\end{align}
and the potential, defined by \eqref{Potential}, now has the form
\begin{equation}
    V(\phi, \chi) = \frac{1}{2}(1-\phi^2 -r\chi^2)^2 + 2\, r^2 \phi^2\,\chi^2.
\end{equation}

If the parameter $r$ is negative, we only have two minima at $(\pm 1, 0)$, which determines only one BPS sector with energy $E_{1,2} = 4/3$. But, if the parameter $r$ is considered to be positive, the model has minima at $v_{1,2} = (\pm 1, 0)$ and $w_{3,4} = (0, \pm 1/\sqrt{r})$. So it has 6 topological sectors: one is a BPS sector with energy $E_{1,2} = 4/3$, four are BPS sectors with degenerate energy $E_{1,3} = E_{1,4} = E_{2,3} = E_{2,4} = 2/3$ and one is a non-BPS sector. To find solutions for this model, we need to search for an orbit that connects two different minima. Thus, to identify this orbit we will apply the integrating factor method on the first-order equations, a procedure outlined in \cite{IntegratingFactor}. The equations of motion in \eqref{EquationofmotionBNRT} could be integrated using the integrating factor $I(\chi) = \chi^{-1 - \frac{1}{r}}$, and that results in
\begin{equation}
    \phi^2 = 1 + \frac{r}{2r-1}\chi^2 + c\chi^{1/r},
\end{equation}
where c is a real constant which determines the shape of the orbit. We will be  concerned with orbits that connect the minima $(\pm 1, 0)$ which have an elliptical form, taking $c = 0$. In this case the orbits are described by
\begin{equation}
\label{OrbitBRNT}
    \phi^2 +\frac{r}{1-2r}\chi^2=1.
\end{equation}

\begin{figure}[!ht]
    \centering
    \includegraphics[scale=0.3]{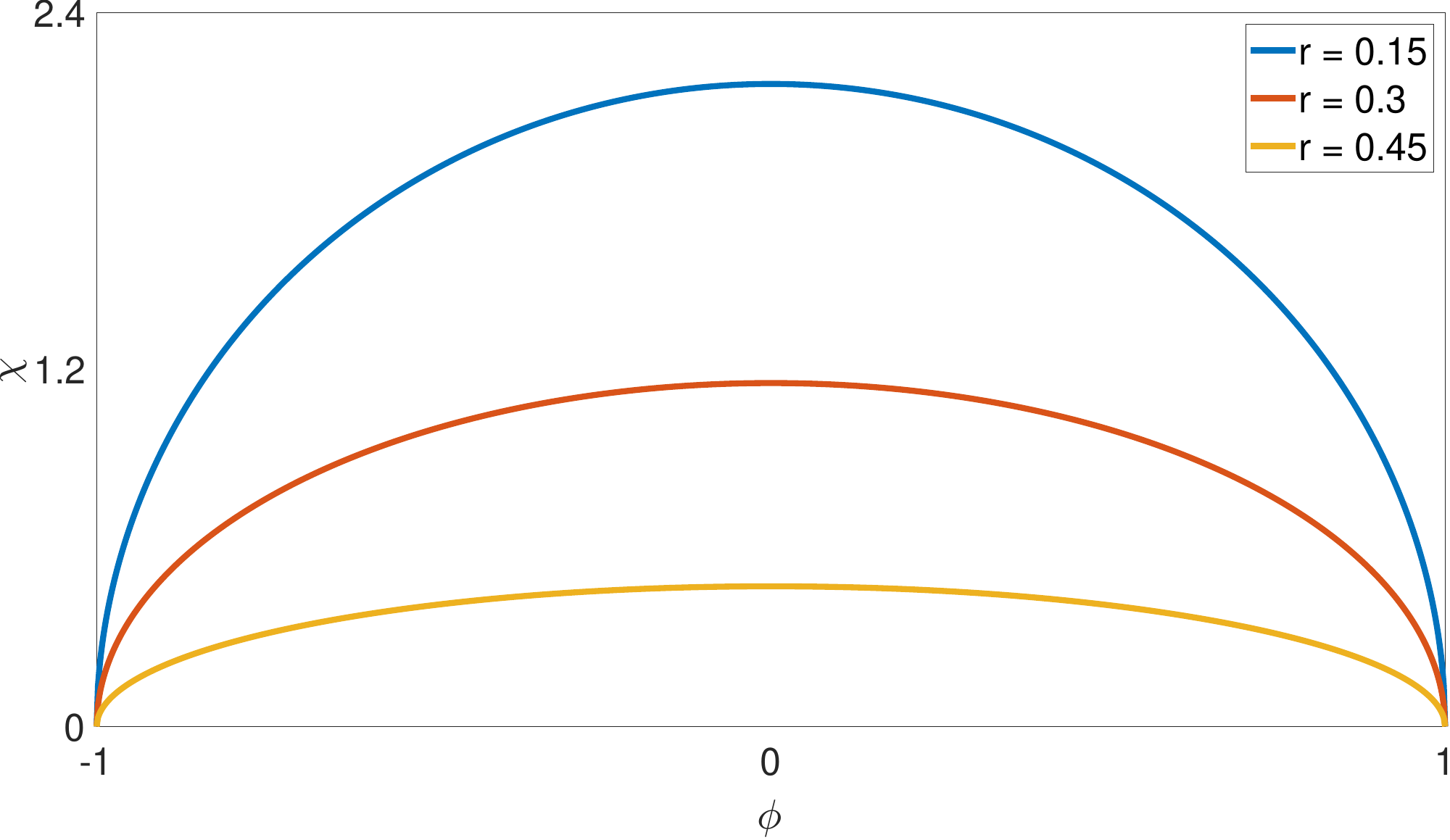}
    \caption{Some orbits for the BNRT model}
    \label{fig:BNRT Orbit}
\end{figure}

\begin{figure}[!ht]
    \centering
    \includegraphics[scale=0.3]{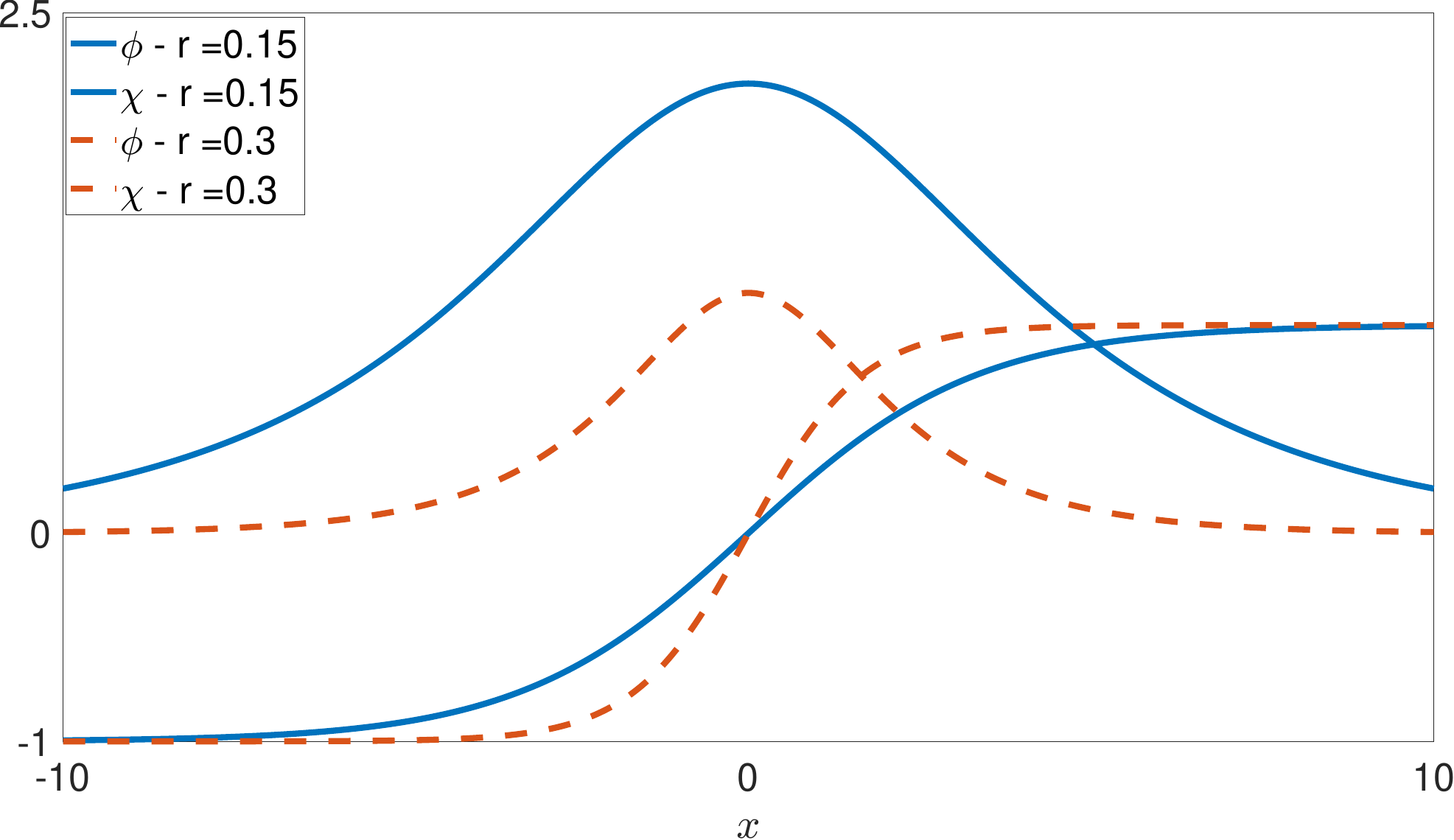}
    \caption{The $\phi$ and $\chi$ solutions for the BNRT model, for $r=0.15$ (blue) and $0.3$ (orange)}
    \label{fig:Solução BNRT}
\end{figure}

We call attention to the fact that when $r\rightarrow {1}/{2}$, the elliptical shape of the orbits are modified into a straight line that connects the pair of minima $(\pm 1,0)$. Here one sees that the parameter $r$ is defined over the interval $r \in (0,1/2)$, and some orbits are depicted in Fig. \ref{fig:BNRT Orbit}. Using the Eq. \eqref{OrbitBRNT} we are able to rewrite the first-order equation for the field $\phi$ as
\begin{equation}
    \phi ' = \pm 2r (1-\phi^2),
\end{equation}
which has the solutions
\begin{equation}
    \phi_{\pm}(x) = \pm \tanh(2rx).
\end{equation}
Substituting this result in the first-order equation \eqref{PrimeiraOrdem} for the field $\chi$, one gets
\begin{equation}
    \chi_{\pm}(x) = \pm \sqrt{\frac{1}{r} - 2}\;\sech(2rx).
\end{equation}
These solutions are depicted in Fig. \ref{fig:Solução BNRT}, for two distinct values of $r$. We notice that both $\phi$ and $\chi$ have the same width $\ell \propto  1/2r$. Also, we can make them to have the same amplitude for $r=1/3$. 

\begin{figure}[!ht]
    \centering    \includegraphics[scale=0.3]{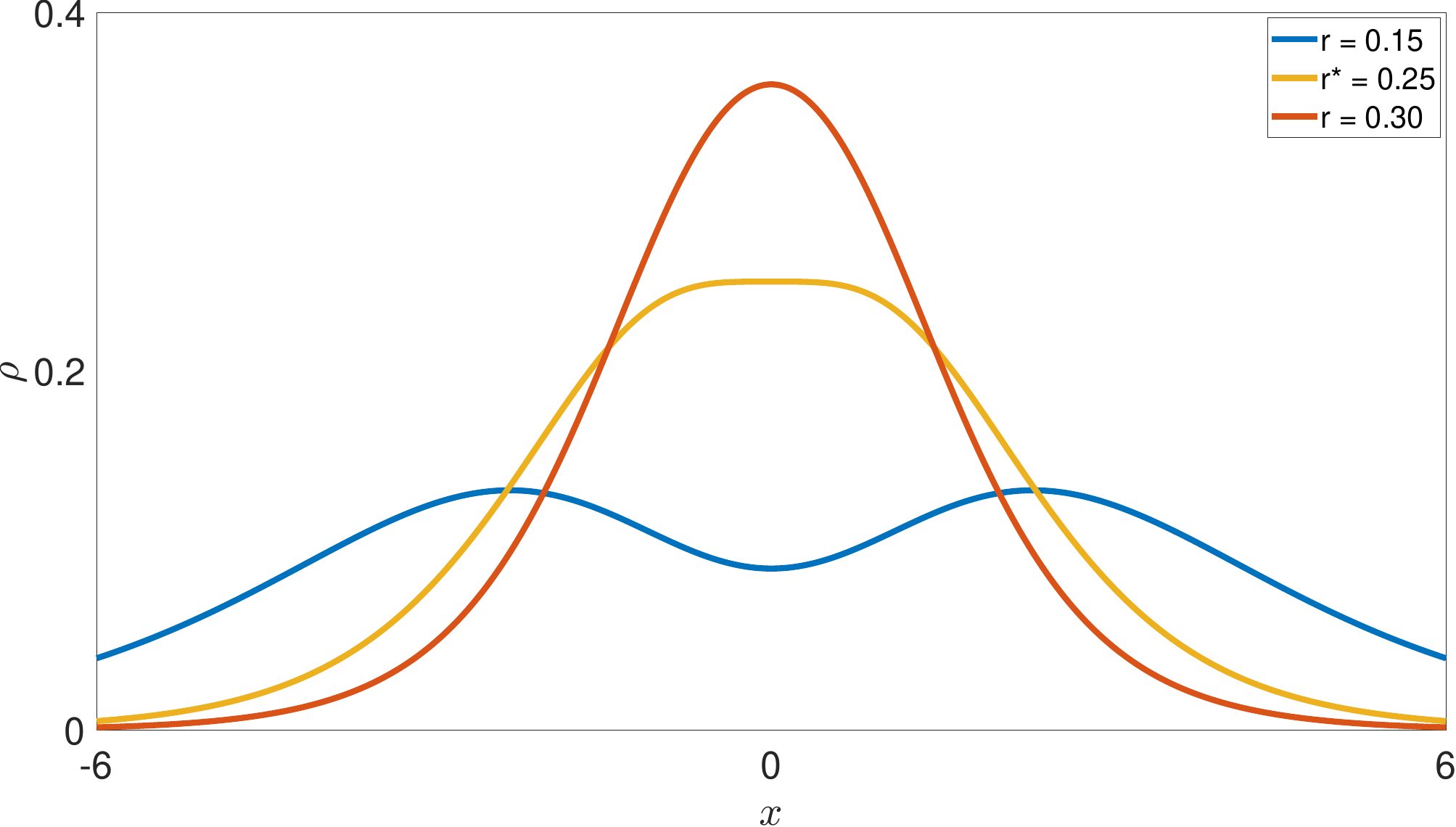}
    \caption{Energy density for the BNRT model.}
    \label{fig:Energy Density for the BNRT Model}
\end{figure}

In this case, the energy density is given explicitly by
\be
    \rho = (1 - \tanh^2(2rx) - (1-2r)\sech^2(2rx))^2 +    
    4(r-2r^2)\tanh^2(2rx)\sech^2(2rx),
\ee
which is depicted in Fig. \ref{fig:Energy Density for the BNRT Model} for some values of $r$. Here we notice that the value $r^*=0.25$ identify the two branches $r\in (0,0.25)$ and $r\in(0.25, 0.5)$, where the energy has two maxima or a single maximum, respectively. This identify the splitting of the energy density, and we further notice that the two fields have the same amplitude for $r=1/3$, so inside the interval $r\in(0.25,0.5)$, in the region outside the splitting of the energy density. 

\subsection{Modified Dynamics}

Let us now describe a model with modified dynamics, by including a function $f=f(\chi)$ of the second field on the kinetic term of $\phi$, as discussed in \cite{Const}. This will also couple the fields and alter the dynamics of them. The model is described by the Lagrangian density 
\begin{equation}\label{modify}
    \mathcal{L} = \frac{1}{2}f(\chi)\partial_{\mu}\phi\partial^{\mu}\phi + \frac{1}{2}\partial_{\mu}\chi\partial^{\mu}\chi - V(\phi, \chi),
\end{equation}
where $f(\chi)$ is a non-negative function of $\chi$ which modifies the kinetic term of the field $\phi$. The equations of motion for $\phi(x,t)$ and $\chi(x,t)$ are now given by 
\begin{align}
\begin{split}
\label{EoMt}
 \partial_{\mu}\PR{f(\chi)\partial^{\mu}\phi} + V_{\phi} = 0,
\;\;\;\;\;\;\;\;
\partial_{\mu}\partial^{\mu}\chi + V_{\chi} - \frac{1}{2}\frac{d f(\chi)}{d \chi}\partial_{\mu}\phi\partial^{\mu}\phi = 0.
\end{split}
\end{align}
Considering static configurations, they change to
\begin{align}
\begin{split}
\label{Second Order - Modified BNRT}
 \frac{d}{dx}\PC{f(\chi)\frac{d\phi}{dx}} = V_{\phi},
\;\;\;\;\;\;\;\;
\frac{d^2\chi}{dx^2} - \frac{1}{2}\frac{df(\chi)}{d\chi}\left({\frac{d\phi}{dx}}\right)^2 = V_\chi.
\end{split}
\end{align}
The energy density can be obtained straightforwardly, giving 
\begin{equation}
    \rho = \frac{1}{2}f(\chi)\PC{\frac{d\phi}{dx}}^2 +\frac{1}{2}\PC{\frac{d\chi}{dx}}^2 + V(\phi,\chi).
\end{equation}

Again, due to the difficulty of working with the second order differential equations \eqref{Second Order - Modified BNRT}, one can use the Bogomol'nyi procedure \cite{Bogomol'nyi} to rewrite the energy density, making use of the auxiliary function $W(\phi,\chi)$
\be
    \rho = \frac{f(\chi)}{2}\PC{\frac{d\phi}{dx} \mp \frac{W_{\phi}}{f(\chi)}}^2 + \frac{1}{2}\PC{\frac{d\chi}{dx} \mp W_{\chi}}^2 +    
    V - \PC{\frac{1}{2}\frac{W_{\phi}^2}{f(\chi)} + \frac{1}{2}W_{\chi}^2} \pm \frac{dW}{dx}.
\ee
Notice that the energy density is minimized if the static fields obey the first-order equations 

\begin{align}
\begin{split}
\label{PrimeiraOrdem2}
 \frac{d\phi}{dx} = \pm \frac{W_{\phi}}{f(\chi)},
\;\;\;\;\;\;\;\;
\frac{d\chi}{dx} = \pm W_{\chi},
\end{split}
\end{align}
when the potential is written as
\begin{equation}
\label{Potencial Dinâmica Modificada}
    V(\phi,\chi) = \frac{1}{2}\frac{W_{\phi}^2}{f(\chi)} + \frac{1}{2}W_{\chi}^2.
\end{equation}
An important remark is that the potential now depends on the function $f(\chi)$. As a consequence, the minima of potential also depends on how the function $f(\chi)$ is defined. Interestingly, the energy lower bound $E_{B}$ is  defined as
\begin{align}
E_{B} = \vert W(\phi (\infty), \chi (\infty)) - W(\phi(-\infty), \chi (-\infty))\vert,
\end{align}
and this reveals that $E_{B}$ just depends on the asymptotic values of the fields, as in the previous case, with standard dynamics.

 Although the solutions are stable due to the BPS procedure, linear stability may be important in the context of quantum corrections and also, for the scattering of kinks. This requires that we add small perturbations around the static configurations, which are given by $\phi(x,t) = \phi(x) + \eta(x)\cos(\omega t)$ and $\chi(x,t) = \chi(x) + \zeta(x)\cos(\omega t)$, supposing that $\eta$ and $\zeta$ are small fluctuations. After substituting these expressions on the equations of motion \eqref{EoMt} we get, up to first order in $\eta(x)$ and $\zeta(x)$,

\begin{equation}\label{LE}
\begin{split}
-\eta_{xx} -\frac{1}{f}f_{\chi}\chi_{x}\eta_{x} + \frac{1}{f}V_{\phi \phi}\eta - \frac{1}{f}f_{\chi}\phi_{x}\zeta_{x} + \frac{1}{f}\PC{V_{\phi\chi} - \phi_{x}\chi_{x}f_{
\chi\chi}-\phi_{xx}f_{\chi}}\zeta &= \omega^2 \eta ,
\\
-\zeta_{xx} + \PC{V_{\chi\chi} + \frac{1}{2}f_{\chi\chi}\phi_{x}^2}\zeta + f_{\chi}\phi_{x}\eta_{x} + V_{\chi\phi}\eta &= \omega^2 \zeta .
\end{split}
\end{equation}
As we can see, these two equations form a complicated set of second order ordinary differential equations, which is in general very hard to solve analytically, even when one considers simple but nontrivial possibilities as the one described in Sec. {\bf B.1} below. 
However, since in this work we are concerned neither with quantum corrections nor with scattering of kinks, we will not deal with solutions to the two equations above. But we can still prove analytically the general result, that the zero mode, which is described by Eq. \eqref{LE} with $\omega=0$, is given by $\eta_0(x)\propto d\phi(x)/dx$, and $\zeta_0(x)\propto d\chi(x)/dx$, when both $\phi(x)$ and $\chi(x)$ solve the first-order equations \eqref{PrimeiraOrdem2}, which is necessary to ensure that the static fields fulfill the Bogomo'nyi bound.
 
{\bf {B.1. The 4-4 model --}} To illustrate the procedure, let us first take the auxiliary function $W(\phi,\chi)$ to have the form
\be
W(\phi,\chi) = \phi - \frac{1}{3}\phi^3 + \alpha\chi - \frac{1}{3}\alpha\chi^3,
\ee
where $\alpha$ is a nonnegative real parameter which controls the intensity of the field $\chi$ and the profile of the solutions. In this case, we have joined the $\phi^4$ model with the $\chi^4$ model, but they do not interact directly, so we call it the 4-4 model. Following the equation \eqref{Potencial Dinâmica Modificada}, the potential for this model is written as
\be
V(\phi,\chi) = \frac{1}{2}\chi^2(1-\phi^2)^2 + \frac{1}{2}\alpha^2(1-\chi^2)^2.
\ee
Here we are using $f(\chi)=1/\chi^2$. It was studied before in Ref. \cite{Const}, so we briefly revise here some of its main features. In particular, the profile of the potential is shown in Fig. \ref{Potencial-(4,4)}, and it has minima at $\phi_\pm=\pm1$ and $\chi_\pm=\pm1$.

\begin{figure}[!ht]
    \centering
    \includegraphics[scale = 0.3]{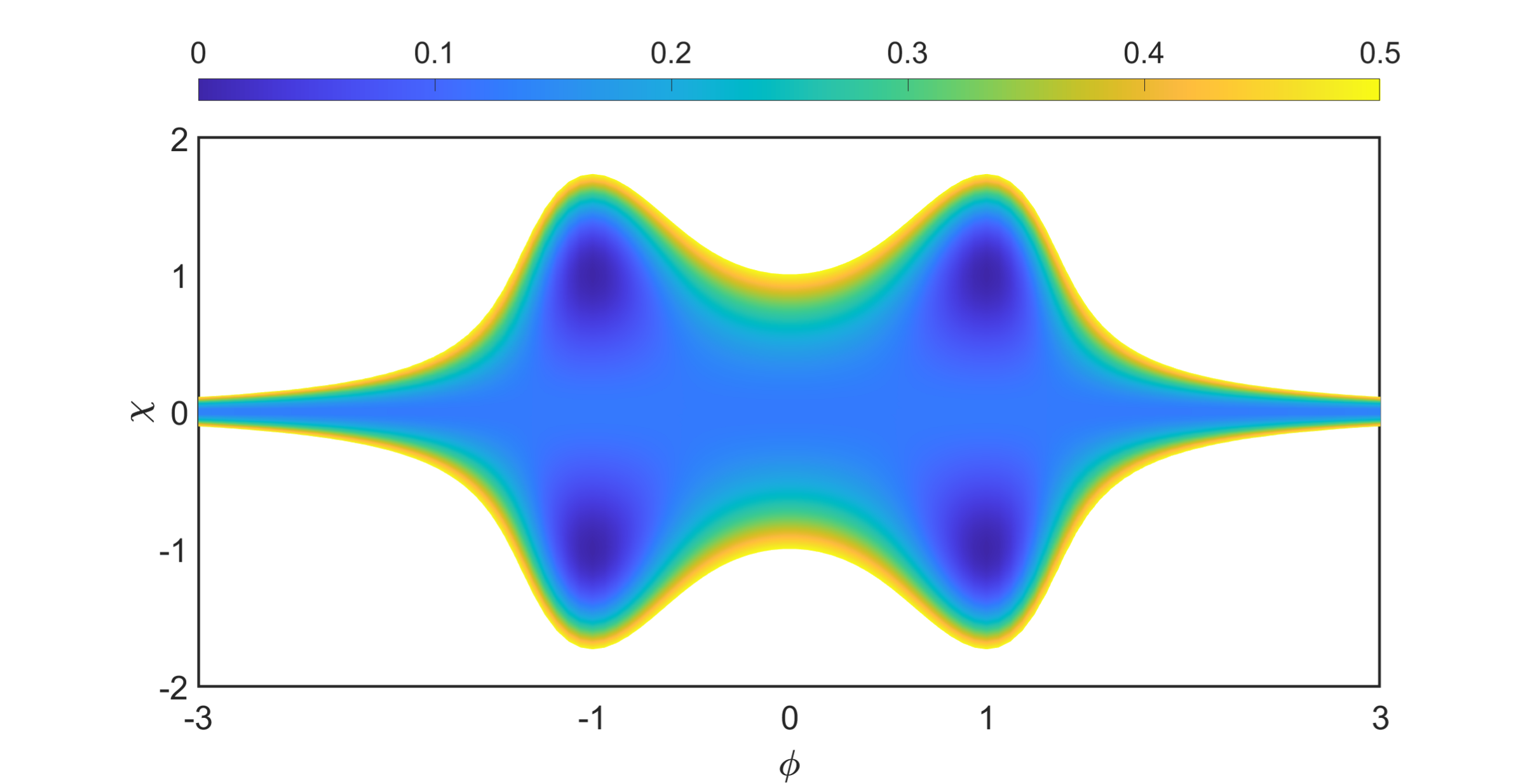}
    \caption{Top view of the potential of the 4-4 model, for $\alpha=0.5$}
    \label{Potencial-(4,4)}
\end{figure}

Since in this model we have $W(\phi,\chi) = W_1(\phi) + W_2(\chi)$, the first order equations become, assuming the plus signals of (21),
\begin{align}
\begin{split}
\label{PrimeiraOrdem26}
 \frac{d\phi}{dx} = \chi^2(1-\phi^2),
\;\;\;\;\;\;\;\;
\frac{d\chi}{dx} = \alpha (1-\chi^2).
\end{split}
\end{align}
As we see, the first-order equation for the field $\chi$ is independent, and can be easily solved, leading to
\be
\chi = \tanh(\alpha x).
\ee
Substituting this result in the equation for the field $\phi$, leads to the solution
\be
\phi = \tanh \left(Y_\alpha(x)\right),
\ee
where $Y_\alpha(x)$ is given by
\be\label{Yx}
Y_\alpha(x)= x - \frac{1}{\alpha}\tanh(\alpha x),
\ee
which responds for the geometric modification included in the model \eqref{modify}. Here one notices that $\chi$ is independent, but it modifies the behavior of the field $\phi$. 
These solutions are depicted in Fig. \ref{Solução - (4,4)} for some values of $\alpha$.
Also, they contribute to the energy density in the form
\be
\rho = \tanh^2(\alpha x)\sech^4\left(Y_\alpha(x)\right) + \alpha^2 \sech^4(\alpha x).
\ee

\begin{figure}[!ht]
    \centering
    \includegraphics[scale=0.3]{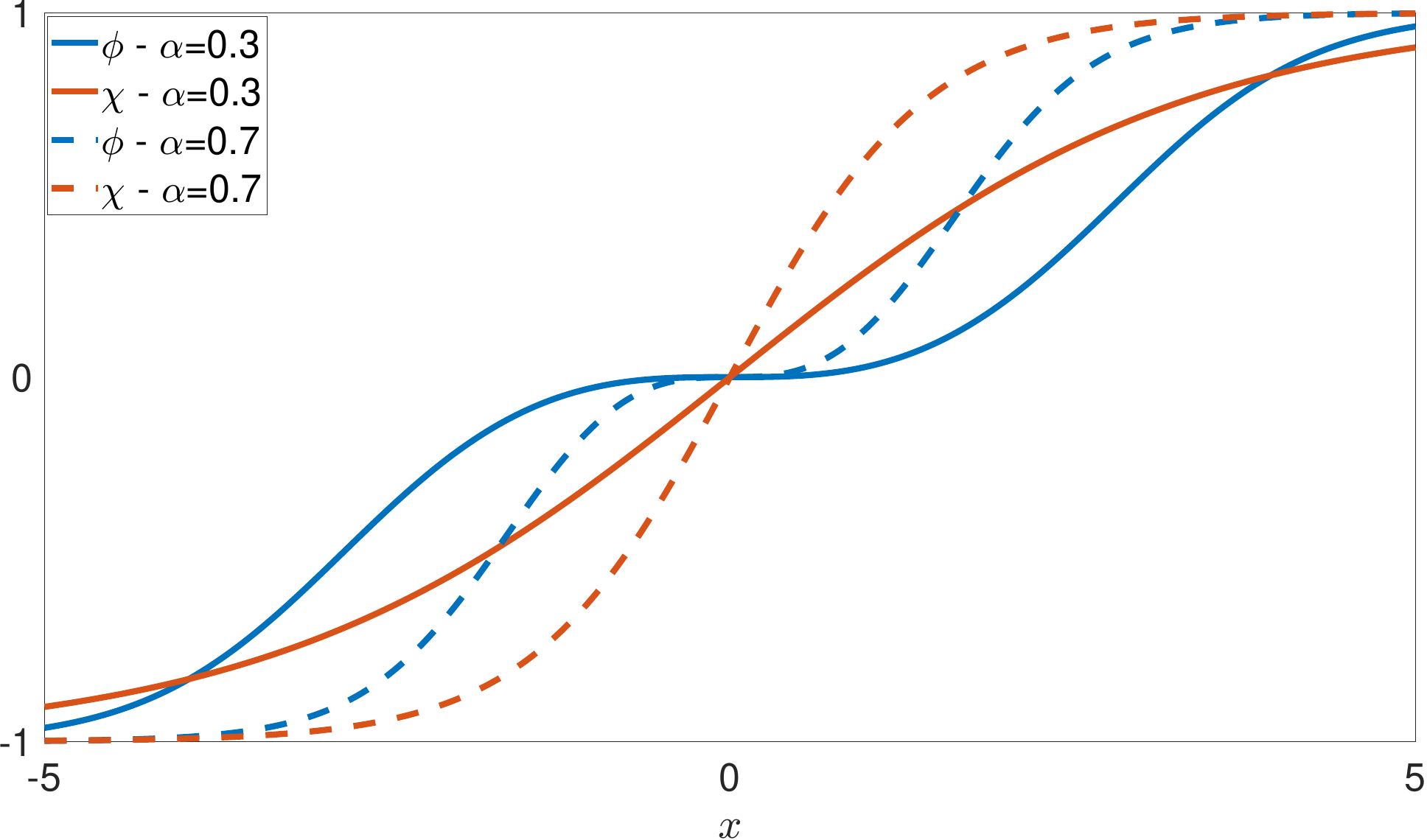}
    \caption{The $\phi$ and $\chi$ solutions of the 4-4 model.}
    \label{Solução - (4,4)}
\end{figure}

Here we want to highlight the fact that $\phi$ exhibits a double kink behavior, engendering a plateau at its center. This profile is due to the kink of the field $\chi$ and the function $f(\chi)$ that modifies the behavior of the field $\phi$. The appearance of the plateau is of current interest, since it is directly related to the presence of the geometric constriction investigated before in Ref. \cite{prbrc}.

\section{New Systems}\label{R}

After briefly describing the main properties of the models with standard and modified dynamics, let us now focus on other systems, to investigate distinct manifestations of the geometric constriction introduced by the presence of the function $f(\chi)$ 
and the $\chi(x)$ configuration itself. We shall consider several possibilities, taking $f(\chi)=1/\chi^2$ for all cases, and using $W(\phi,\chi)=W_1(\phi)+W_2(\chi)$. This means that $\chi$ acts independently, that is, it is not modified by the behavior of the other field, $\phi$. 

\subsection{The 4-6 model}

Here we want to study how the scalar field $\chi$ contributes to change the profile of the field $\phi$, when it engenders the sixth order power contribution. To make this possibility concrete, we take the auxiliary function $W(\phi,\chi)$ to be of the form
\be
W(\phi,\chi) = \phi - \frac{1}{3}\phi^3 + \frac{1}{2}\alpha\chi^2 - \frac{1}{4}\alpha\chi^4,
\ee
where $\alpha$ is a nonnegative real parameter which controls the intensity of the field $\chi$ and the profile of the solutions. This is the 4-6 model, which puts together the $\phi^4$ model and the $\chi^6$ model. Following the equation \eqref{Potencial Dinâmica Modificada}, the potential for this model is written as
\be
V(\phi,\chi) = \frac{1}{2}\chi^2(1-\phi^2)^2 + \frac{1}{2}\alpha^2\chi^2(1-\chi^2)^2.
\ee
It has a continuum line of minima at $\chi = 0$ and $\phi$ arbitrary. It also has minima at $\phi_\pm=\pm1$ and $\chi_\pm=\pm1$. The profile of this potential is shown in Fig. \ref{Potencial-(4,6)}.

\begin{figure}[!ht]
    \centering
    \includegraphics[scale = 0.3]{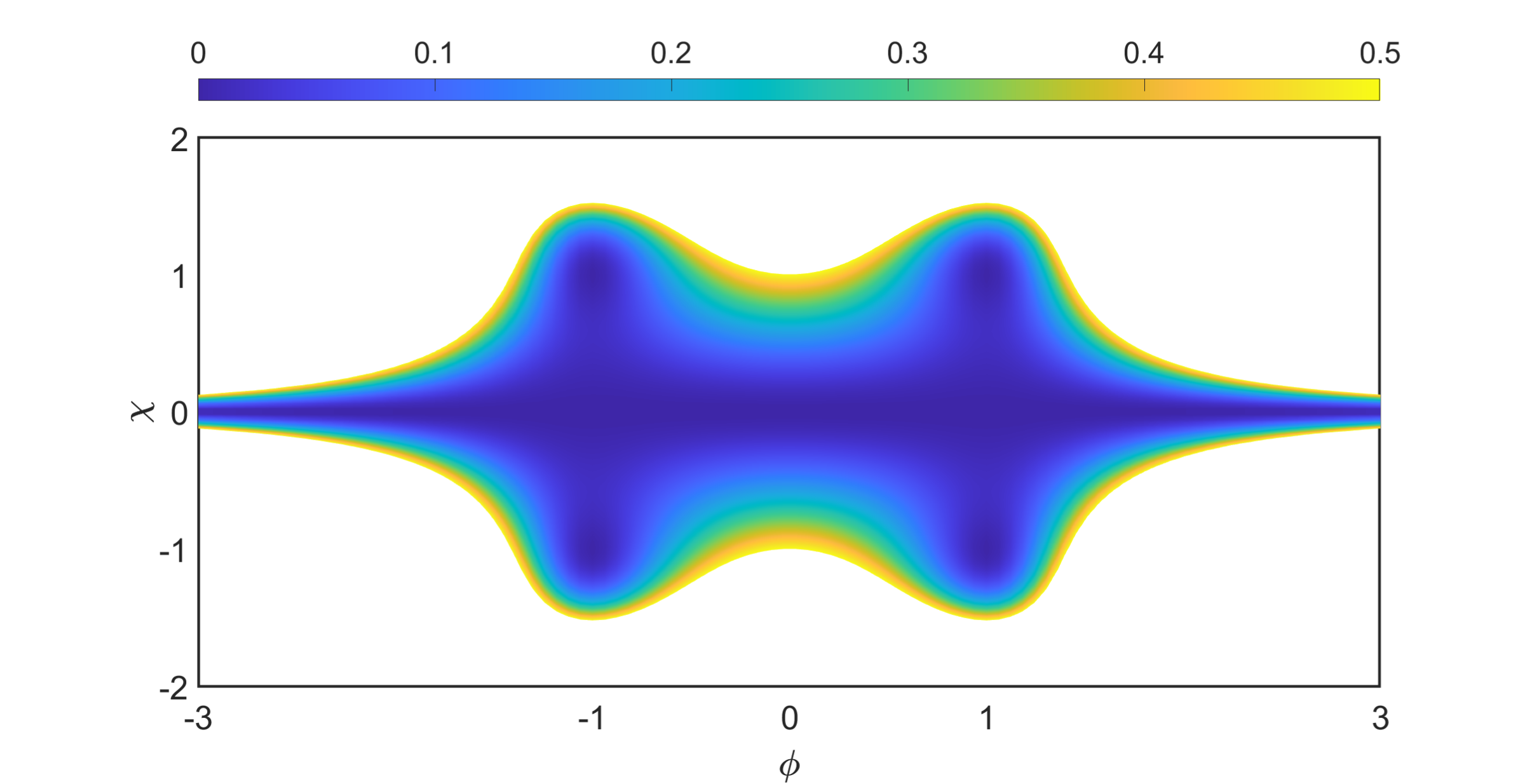}
    \caption{Top view of the potential of the 4-6 model, for $\alpha=0.5$}
    \label{Potencial-(4,6)}
\end{figure}

In this model, the first-order equations \eqref{PrimeiraOrdem2} become (taking the upper signal)
\begin{align}
\begin{split}
\label{PrimeiraOrdem25}
 \frac{d\phi}{dx} = \chi^2(1-\phi^2),
\;\;\;\;\;\;\;\;
\frac{d\chi}{dx} = \alpha \chi(1-\chi^2).
\end{split}
\end{align}
We notice that the equation for the field $\chi$ can be solved independently, and has the solution
\be
\chi(x) = \sqrt{\frac{1}{2}(1+\tanh(\alpha x))}.
\ee
Substituting this result in the equation for the field $\phi$ leads to the solution
\be
\phi(x) = \tanh\left(Z_\alpha(x)\right),
\ee
where
\be\label{Zx}
Z_\alpha(x)=\frac{1}{2}\PC{x+\frac{1}{\alpha}\ln\PC{\cosh(\alpha x)}},
\ee
which responds for the geometric constriction included in the model. It is different from $Y_\alpha(x)$ which appeared in Eq. \eqref{Yx}, since now we are considering the $\chi^6$ model, and this modify the solution $\chi(x)$ and the form of the constraint, as a consequence.

The above solutions are depicted in Fig. \ref{Solução - (4,6)}. Also, the corresponding energy density is given by
\be
\rho = \frac{1}{2}\PC{1 + \tanh(\alpha x)}\sech^4\left(Z_\alpha(x)\right)+ \frac{\alpha^2}{8}\PC{1 + \tanh(\alpha x)}\PC{1 - \tanh(\alpha x)}^2.
\ee

\begin{figure}[!ht]
    \centering
    \includegraphics[scale=0.3]{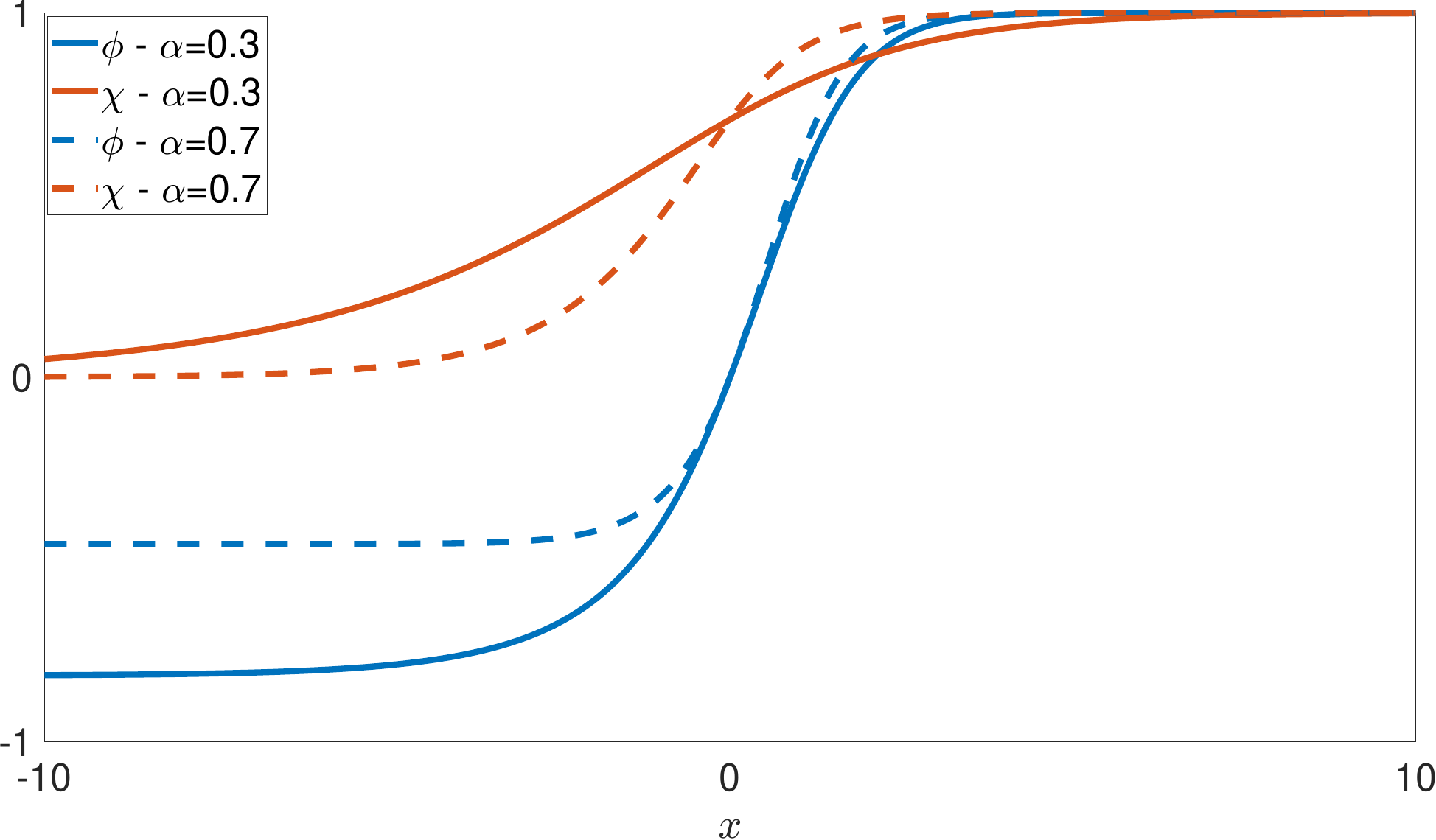}
    \caption{The $\phi$ and $\chi$ solutions of the 4-6 model}
    \label{Solução - (4,6)}
\end{figure}

Here we notice that, differently from the previous example, the solution for the  field $\phi$ doesn't show a plateau at its center. Instead, it has an asymmetry, going to a negative value between $0$ and $-1$, but different from $-1$. This is a consequence of the field $\chi$ only going to zero in the limit where $ x \rightarrow -\infty$. We then see that the $\chi^6$ solution induces an asymmetry in the $\phi^4$ solution, making its left tail to change importantly. This may be of practical use, if one wants to investigate the magnetization in magnetic materials, an issue to be considered elsewhere.

\subsection{The 6-4 model}

In this case we take the auxiliary function $W(\phi,\chi)$ to be
\be
W(\phi,\chi) = \frac{1}{2}\phi^2 - \frac{1}{4}\phi^4 + \alpha\chi - \frac{1}{3}\alpha\chi^3,
\ee
where $\alpha$ is a nonnegative real parameter which acts as in the previous model. Following the equation \eqref{Potencial Dinâmica Modificada}, the potential for this model is written as
\be
V(\phi,\chi) = \frac{1}{2}\chi^2\phi^2(1-\phi^2)^2 + \frac{1}{2}\alpha^2(1-\chi^2)^2.
\ee
This potential has minima at $\phi=0$ and $\chi_\pm=\pm1$, and also at $\phi_\pm=\pm1$ and $\chi_\pm=\pm1$.
The profile of the potential is shown in Fig. \ref{Potencial-(6,4)}.

\begin{figure}[!ht]
    \centering
    \includegraphics[scale = 0.3]{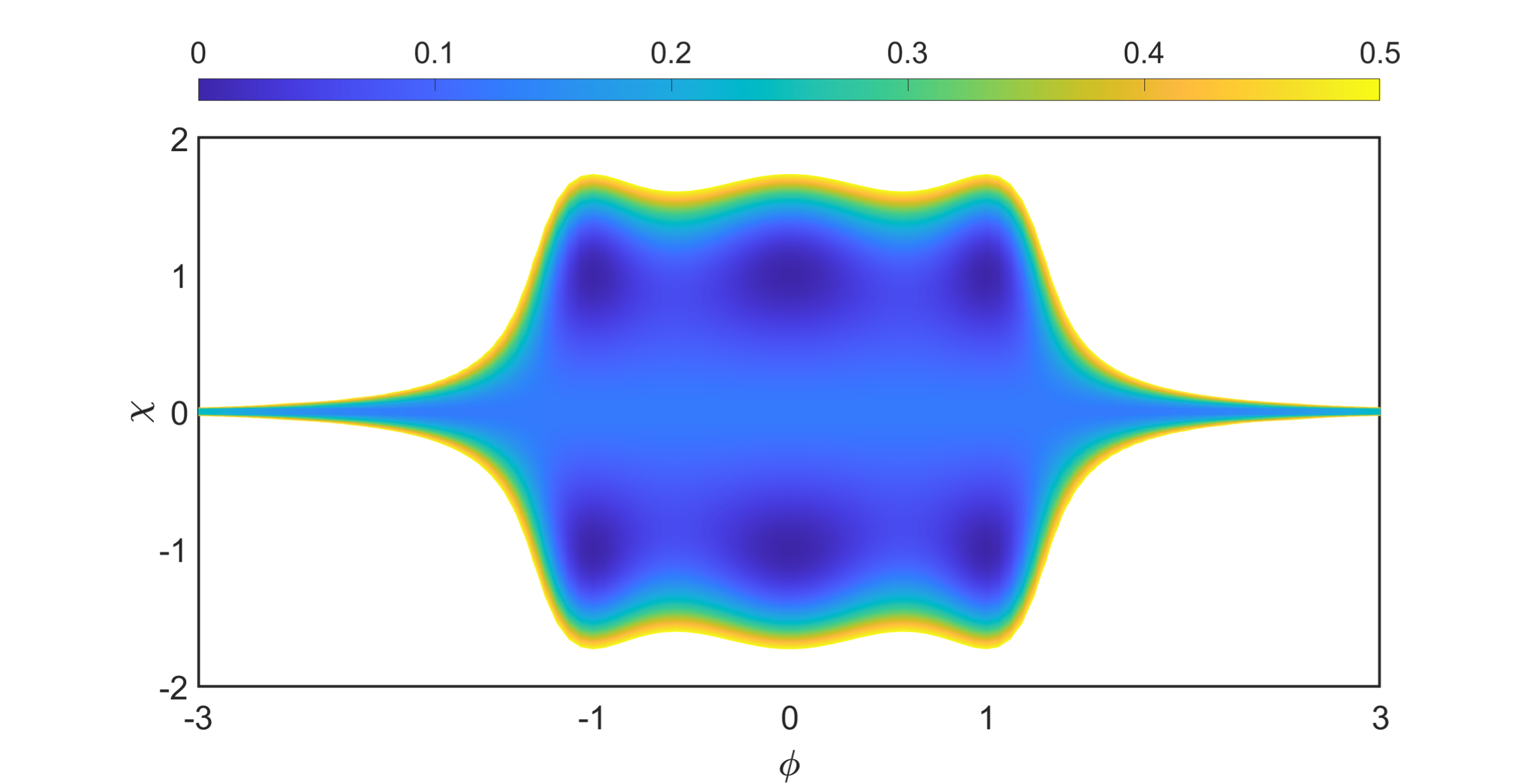}
    \caption{Top view of the potential of the 6-4 model, for $\alpha=0.5$}
    \label{Potencial-(6,4)}
\end{figure}

In this model, the first-order equations \eqref{PrimeiraOrdem2} become, taking the plus signals, 
\begin{align}
\begin{split}
\label{PrimeiraOrdem24}
 \frac{d\phi}{dx} = \chi^2\phi(1-\phi^2),
\;\;\;\;\;\;\;\;
\frac{d\chi}{dx} = \alpha (1-\chi^2).
\end{split}
\end{align}
The equation for the field $\chi$ has the solution
\be
\chi = \tanh(\alpha x).
\ee
Substituting this result in the equation for the field $\phi$ leads to the solution
\be
\phi = \sqrt{\frac{1}{2}\PC{1 + \tanh\PC{Y_\alpha(x)}}},
\ee
where $Y_\alpha(x)$ was already defined in Eq. \eqref{Yx}. These solutions are depicted in Fig. \ref{Solução - (6,4)} for some values of $\alpha$. 
Also, the energy density is given by

\be
\rho = \frac{1}{8}\tanh^2(\alpha x)\PR{1 + \tanh\PC{Y_\alpha(x)}}\PR{1 - \tanh\PC{Y_\alpha(x)}}^2 + \alpha^2\sech^4(\alpha x).
\ee

We notice that the solution of $\phi$ connects the asymmetric minima $0$ and $1$, a characteristic behavior of the kink of the $\phi^6$ model. However, in the present model it engenders a plateau around $x=0$, due to the field $\chi$, and the function $f(\chi)$ as well.

\begin{figure}[!ht]
    \centering
    \includegraphics[scale=0.3]{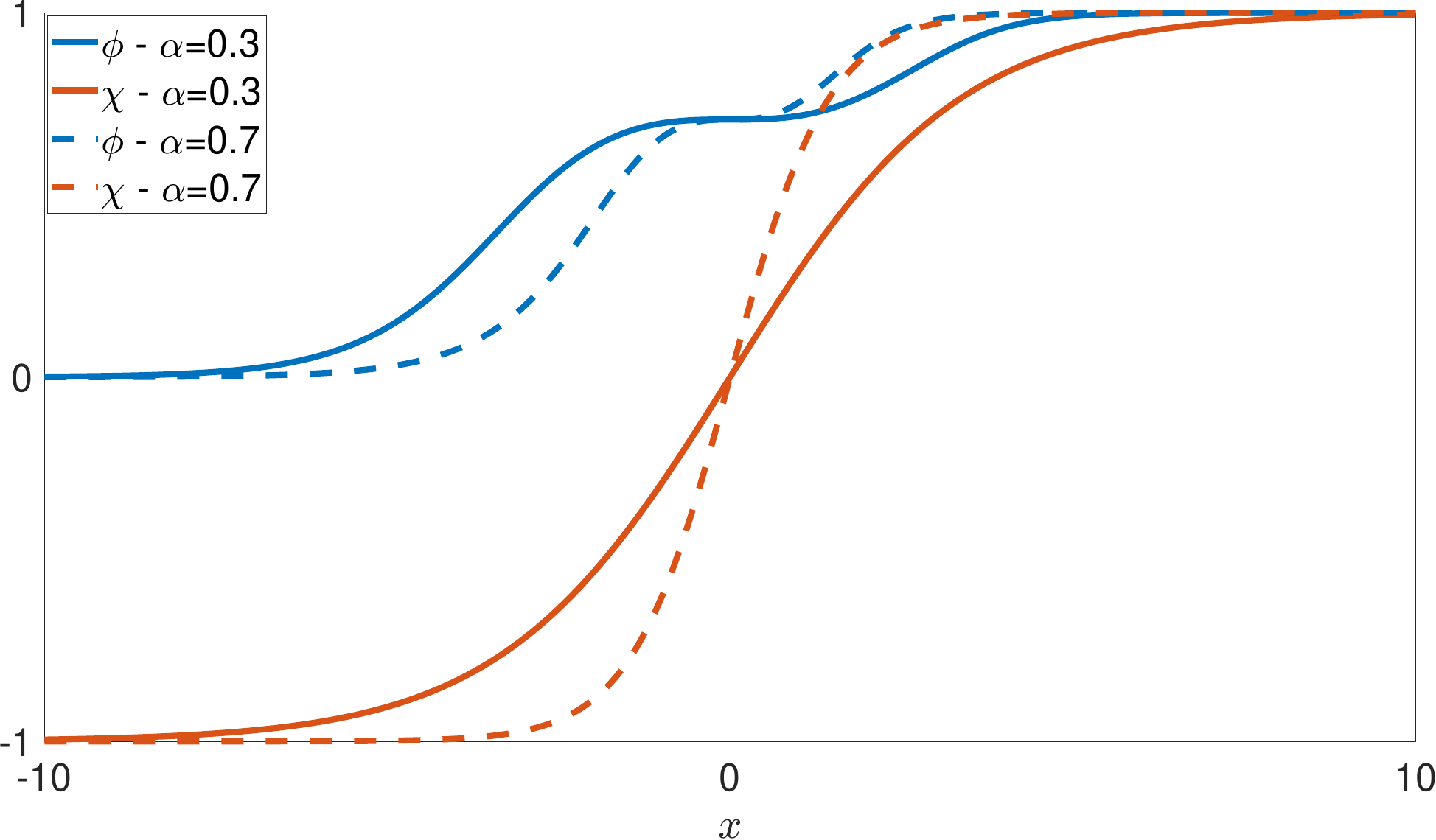}
    \caption{The $\phi$ and $\chi$ solutions for the 6-4 model.}
    \label{Solução - (6,4)}
\end{figure}

\subsection{The 6-6 model}

Let us now consider that the auxiliary function $W(\phi,\chi)$ in the form
\be
W(\phi,\chi) = \frac{1}{2}\phi^2 - \frac{1}{4}\phi^4 + \frac{1}{2}\alpha\chi^2 - \frac{1}{4}\alpha\chi^4,
\ee
where $\alpha$ is a nonnegative real parameter which also acts as in the previous case. Here we are joining the $\phi^6$ model with the $\chi^6$ model, so we refer to it as the 6-6 model. Using the equation \eqref{Potencial Dinâmica Modificada}, the potential is now given by
\be
V(\phi,\chi) = \frac{1}{2}\chi^2\phi^2(1-\phi^2)^2 + \frac{1}{2}\alpha^2\chi^2(1-\chi^2)^2.
\ee
This potential has several minima, in particular, the continuum of minima at $\chi=0$ and $\phi$ arbitrary; see Fig. \ref{Potencial-(6,6)}. Here, the first-order equations become, after taking the plus signals in \eqref{PrimeiraOrdem2},
\begin{align}
\begin{split}
\label{PrimeiraOrdem23}
 \frac{d\phi}{dx} = \chi^2\phi(1-\phi^2),
\;\;\;\;\;\;\;\;
\frac{d\chi}{dx} = \alpha \chi(1-\chi^2).
\end{split}
\end{align}
We notice that the equation for the field $\chi$ can be easily solved to give 
\be
\chi = \sqrt{\frac{1}{2}(1+\tanh(\alpha x))},
\ee
such that the equation for the field $\phi$ includes the solution
\be
\phi = \sqrt{\frac{1}{2}\PC{1 + \tanh\PC{Z_\alpha(x)}}},
\ee
where $Z_\alpha(x)$ was already defined in Eq. \eqref{Zx}.

\begin{figure}[!ht]
    \centering
    \includegraphics[scale = 0.3]{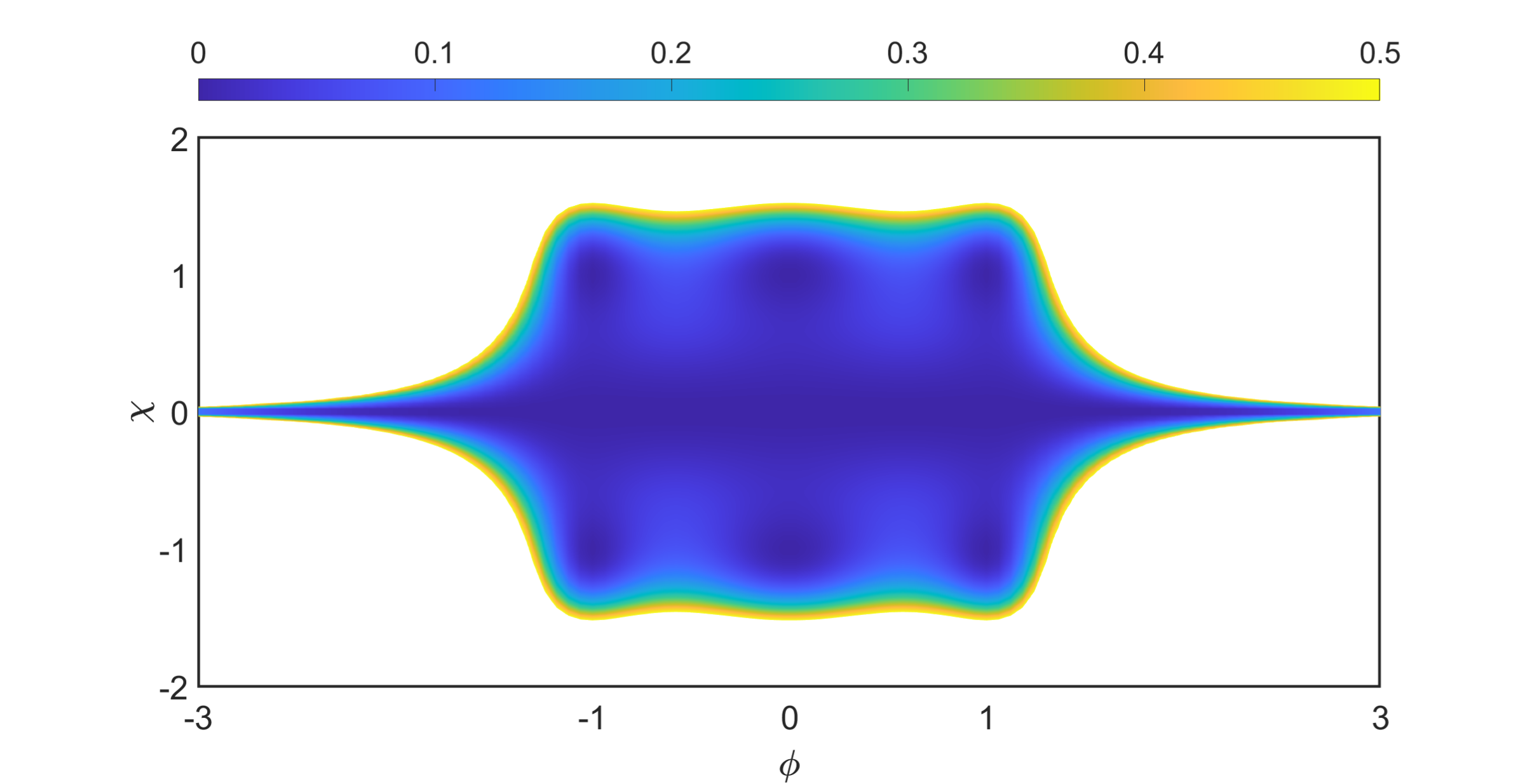}
    \caption{Top view of the potential of the 6-6 model, for $\alpha=0.5$}
    \label{Potencial-(6,6)}
\end{figure}

These solutions can be seen in Fig. \ref{Solução - (6,6)}. We see a shift in the minimum connected by the left tail, which is not more bounded by the minimum at zero as seen in the standard $\chi^6$ model. This shows that the $\chi$ field here changes the left tail of the field $\phi$ in an important way, giving room for application of current interest.
\begin{figure}[!ht]
    \centering
    \includegraphics[scale=0.3]{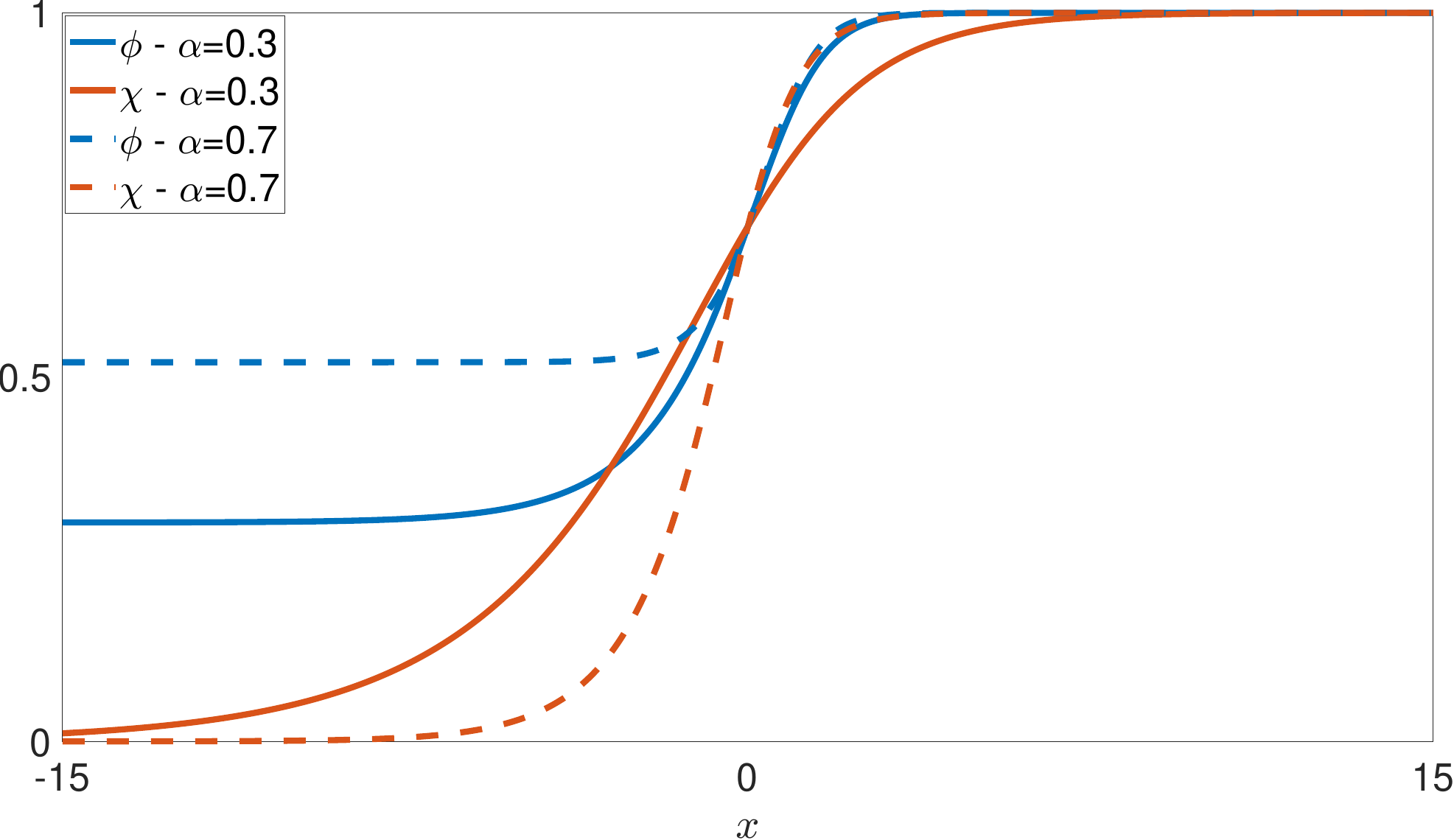}
    \caption{The $\phi$ and $\chi$ solutions for the 6-6 model.}
    \label{Solução - (6,6)}
\end{figure}

This model has energy density given by
\begin{align}
\begin{split}
\rho = \frac{1}{16}(1+\tanh(\alpha x))\PC{1 + \tanh\PC{Z_\alpha(x)}}\PC{1 - \tanh\PC{Z_\alpha(x)}}^2 +
\frac{\alpha^2}{8}(1 + \tanh(\alpha x))\PC{1 - \tanh(\alpha x)}^2.
\end{split}
\end{align}

\subsection{The V-4 model}

Another example to be investigated is when the field $\phi$ is defined following the lines of Refs. \cite{Vil,Vacuumless}. In \cite{Vil}, the potential was called vacuumless, so we use V to refer to it in this work, since the potential has a local maximum at $\phi = 0$ and minima only when $\phi\to \pm\infty$. Here we take the auxiliary function $W(\phi,\chi)$ to be given by
\be
W(\phi, \chi) = \arctan[\sinh(\phi)] + \alpha\chi - \frac{1}{3}\alpha\chi^3,
\ee
where $\alpha$ is a nonnegative real parameter which acts to control the intensity of the field $\chi$ and the profile of the solutions. This defines the V-4 model. Following the equation \eqref{Potencial Dinâmica Modificada}, the potential of the model can be written as
\be
V(\phi, \chi) = \frac{1}{2}\chi^2\,\sech^2(\phi) + \frac{1}{2}\alpha^2(1-\chi^2)^2.
\ee
The profile of the potential is shown in Fig. \ref{Potencial-(Vac,4)}, and the 
first-order equations \eqref{PrimeiraOrdem2} now become, using the plus signals,

\begin{align}
\begin{split}
\label{PrimeiraOrdem22}
 \frac{d\phi}{dx} = \chi^2\sech(\phi),
\;\;\;\;\;\;\;\;
\frac{d\chi}{dx} = \alpha (1-\chi^2).
\end{split}
\end{align}

\begin{figure}[!ht]
    \centering
    \includegraphics[scale = 0.3]{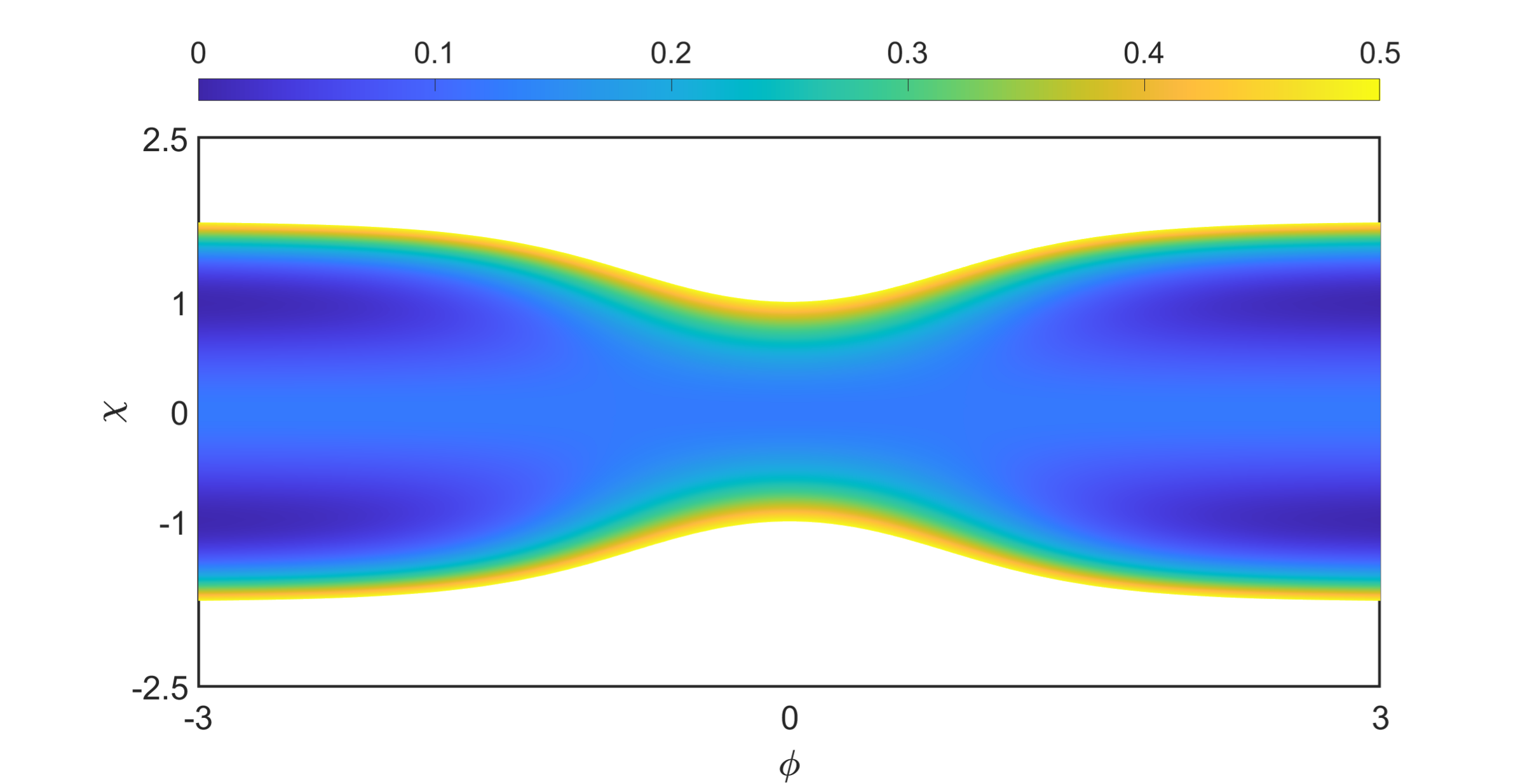}
    \caption{Top view of the potential for the V-4 model, for $\alpha=0.5$}
    \label{Potencial-(Vac,4)}
\end{figure}
The solutions are
\be
\chi = \tanh(\alpha x),
\ee
and 
\be
\phi = \arcsinh \PC{Y_\alpha(x)}.
\ee
This solution presents a plateau around $x=0$, but keeps the vaccumless behavior as we can see in Fig. \ref{Solução - (Vac,4)}. Also, the energy density is given by
\be
\rho = \tanh^2(\alpha x)\chav{\sech\PC{\arcsinh\PC{Y_\alpha(x)}}}^2 +\frac{\alpha^2}{2}\sech^4(\alpha x).
\ee

\begin{figure}[!ht]
    \centering
    \includegraphics[scale=0.3]{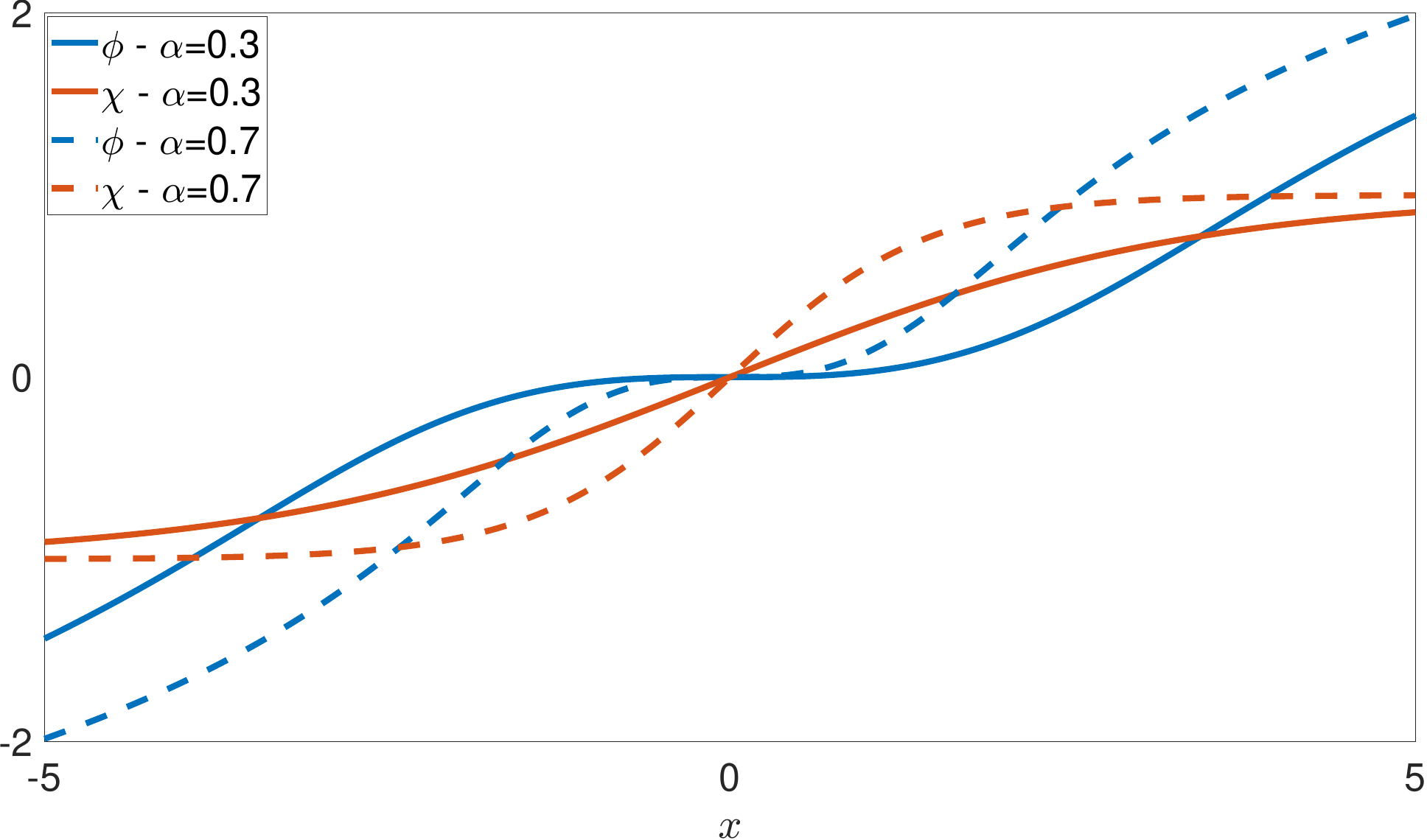}    \caption{The $\phi$ and $\chi$ solutions of  the V-4 model}
    \label{Solução - (Vac,4)}
\end{figure}

\subsection{The V-6 model}

We can also investigate what happens to the vaccumless model, when one considers the other field to be defined as in the $\chi^6$ model. The auxiliary function in this case is given by
\be
W(\phi, \chi) = \arctan[\sinh(\phi)] + \frac{1}{2}\alpha\chi^2 - \frac{1}{4}\alpha\chi^4,
\ee
where $\alpha$ is a nonnegative real parameter which acts as in the previous model. Following the equation \eqref{Potencial Dinâmica Modificada}, the potential has now the form 
\be
V(\phi, \chi) = \frac{1}{2}\chi^2\sech^2(\phi) + \frac{1}{2}\alpha^2\chi^2(1-\chi^2)^2.
\ee
The profile of the potential is shown in Fig. \ref{Potencial-(Vac,6)}, and the first-order equations \eqref{PrimeiraOrdem2} become, taking the plus signals,

\begin{figure}[!ht]
    \centering
    \includegraphics[scale = 0.3]{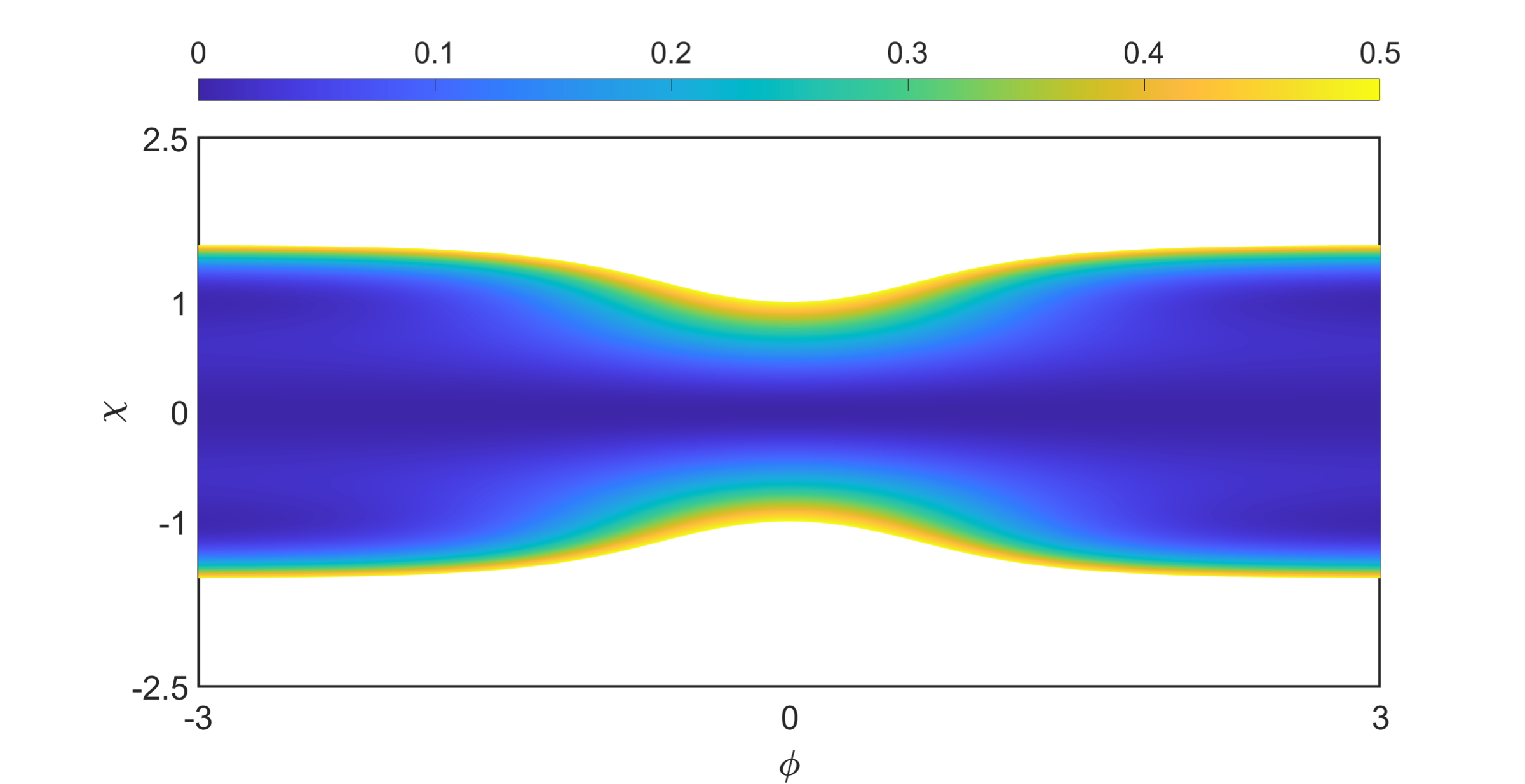}
    \caption{Top view of the potential of the V-6 model, for $\alpha=0.5$}
    \label{Potencial-(Vac,6)}
\end{figure}

\begin{align}
\begin{split}
\label{PrimeiraOrdem21}
 \frac{d\phi}{dx} = \chi^2\sech(\phi),
\;\;\;\;\;\;\;\;
\frac{d\chi}{dx} = \alpha \chi(1-\chi^2).
\end{split}
\end{align}
The field $\chi$ has the solution
\be
\chi = \sqrt{\frac{1}{2}(1+\tanh(\alpha x))},
\ee
and the field $\phi$ behaves as 
\be
\phi = \arcsinh\PC{Z_\alpha(x)}.
\ee

Here we see that the kink of $\chi^6$ model acts changing the behavior of the kink of the vaccumless model in the left tail. Thus, we have a new feature: the left tail of the kink ends at a value in between $0$ and $-1$, the precise value depending on $\alpha$. This is depicted in Fig. \ref{Solução - (Vac,6)}. The energy density is given by

\be
\rho = \frac{1}{2}(1+\tanh(\alpha x))\sech^2\PC{\arcsinh\PC{Z_\alpha(x)}} +\frac{\alpha^2}{8}\PC{1 + \tanh(\alpha x)}\PC{1 - \tanh(\alpha x)}^2.
\ee

\begin{figure}[!ht]
    \centering
    \includegraphics[scale=0.3]{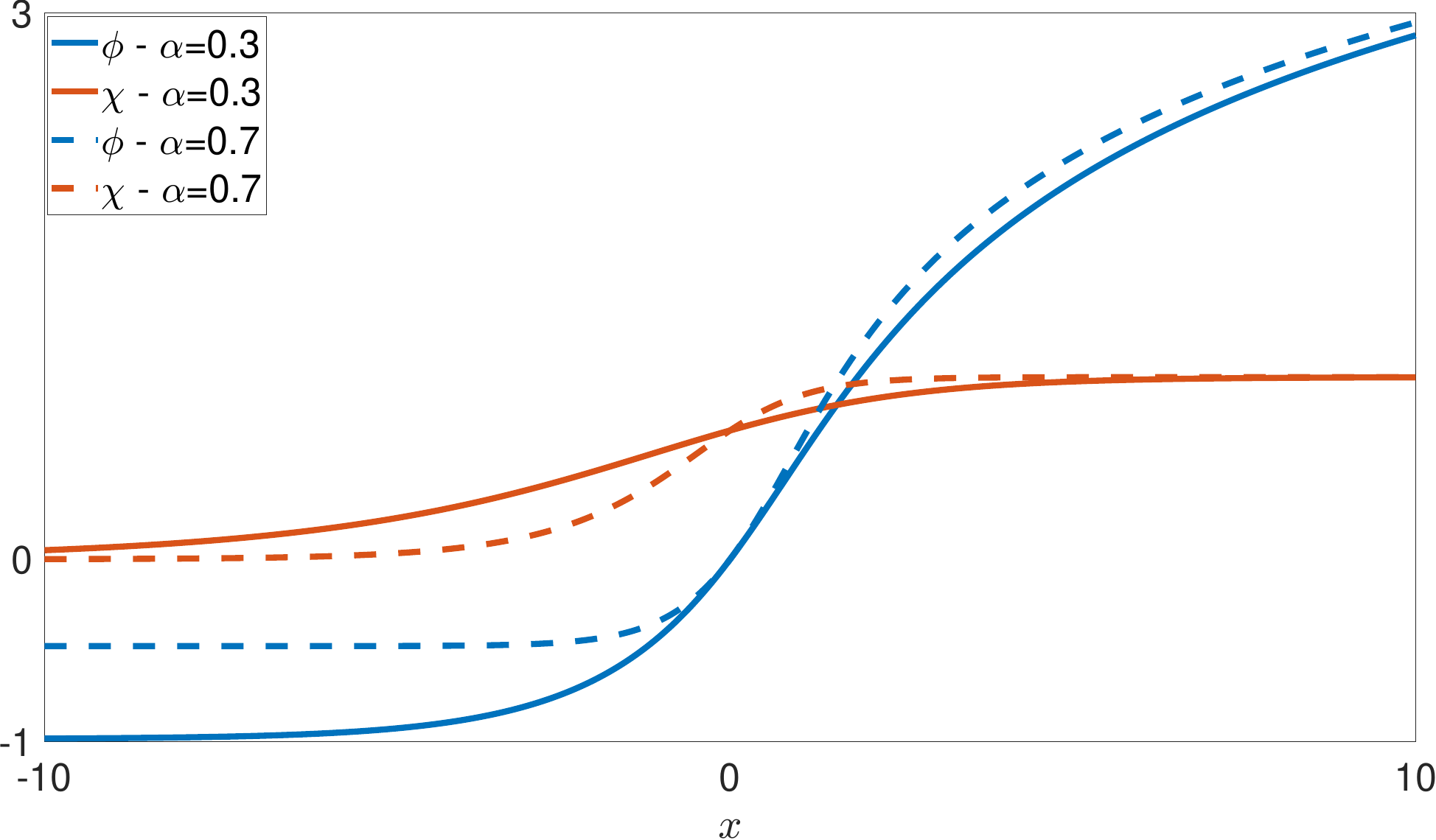}
    \caption{The $\phi$ and $\chi$ solutions for the V-6 model}
    \label{Solução - (Vac,6)}
\end{figure}

\section{Modified BNRT model}\label{IV}

\begin{figure}[!ht]
    \centering \includegraphics[scale = 0.3]{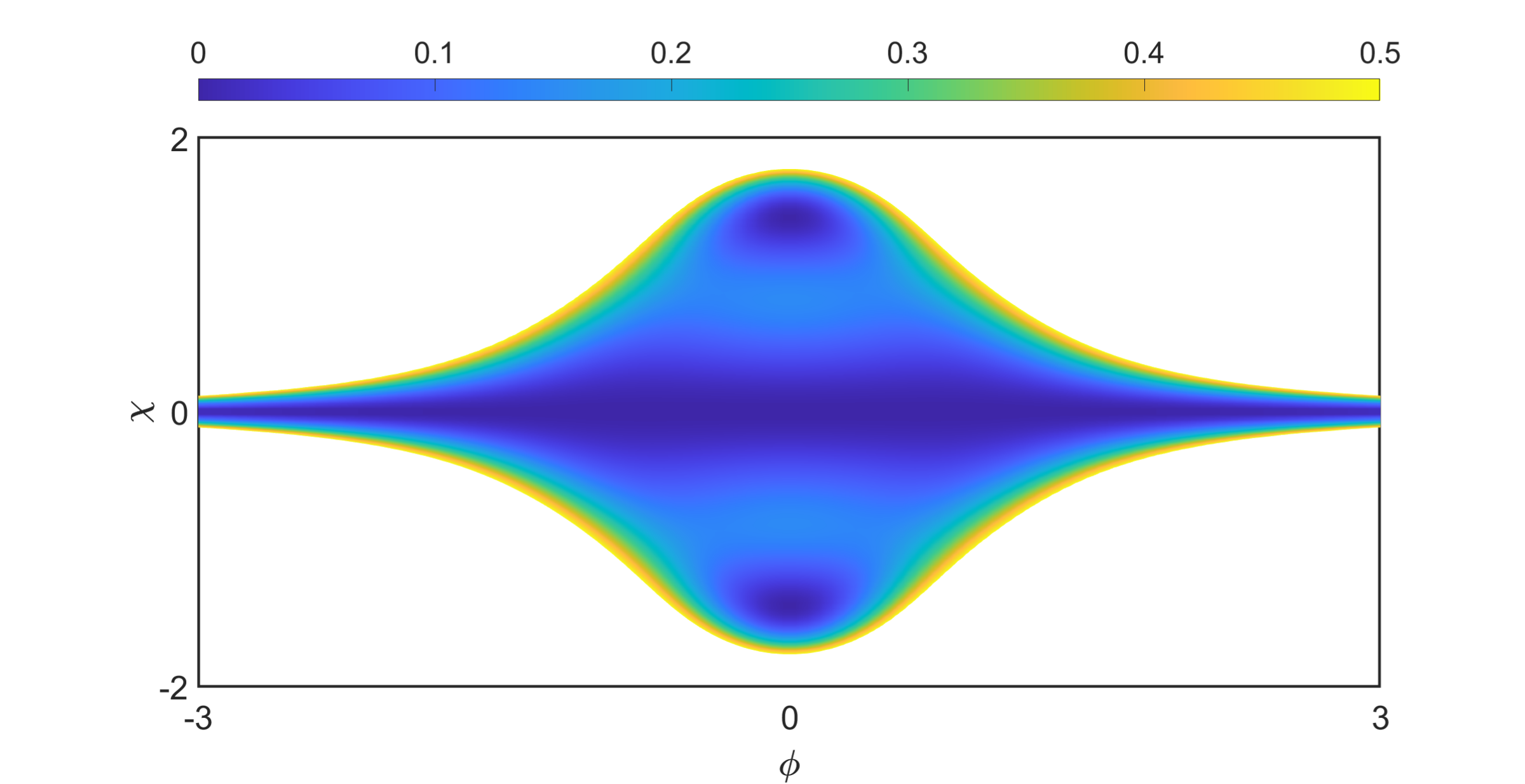}
    \caption{Top view of the potential for the modified BNRT model, for $r=0.5$}    \label{fig:Potential for the Modified BNRT Model}
\end{figure}

Let us now consider a model in which the auxiliary function is not separable anymore. This means that the behavior of the field $\chi$ is now also modified by the other field, $\phi$. In this case, the two first-order equations are coupled, and the problem of finding analytical solutions is much harder than before. However, due to the presence of analytical solutions in the BNRT model, we think we can take a step forward and investigate the modified BNRT model. It is defined by Eq. \eqref{modify}, with $W$ as in Eq. \eqref{BRNT_W}, and we use that same    $f(\chi)=1/\chi^2$. We consider static fields and the equations of motion \eqref{Second Order - Modified BNRT} now become
\begin{align}
\begin{split}
 \frac{d}{dx}\PC{\frac{1}{\chi^2}\frac{d\phi}{dx}} =
 V_{\phi},
\;\;\;\;\;\;\;\;
\frac{d^2\chi}{dx^2} + \frac{1}{\chi^3}\PC{\frac{d\phi}{dx}}^2 = V_\chi.
\end{split}
\end{align}
With $W$ as in Eq. \eqref{BRNT_W}, the first-order equations now become 
\begin{align}
\begin{split}
\label{1OMBNRT}
 \frac{d\phi}{dx} = \chi^2 - \chi^2\phi^2 -r\chi^4,
\;\;\;\;\;\;\;\;
\frac{d\chi}{dx} = -2r\phi\chi.
\end{split}
\end{align}
The potential is written as
\begin{equation}
    V(\phi, \chi) = \frac{1}{2}\chi^2(1-\phi^2 -r\chi^2)^2 + 2r^2\phi^2\chi^2.
\end{equation}
 We notice the presence of minima for $\chi=0$ and $\phi$ arbitrary, and two other minima for $\phi=0$ and $\chi=\pm1/\sqrt{r}$. This potential is shown in Fig. \ref{fig:Potential for the Modified BNRT Model}, for $r=0.5$.

Differently from the standard BNRT model, here a new set on minima appears; it refers to the case of $\chi=0$ and $\phi$ arbitrary, and this will bring new results. In order to solve the first-order equations we need to find an orbit that connects two minima, to accomplish that we use again the method of integrating factor \cite{IntegratingFactor} to get the following orbit
\begin{equation}
    \phi^2 = 1 - 2r^2 - r\chi^2 + \frac{c}{r}e^{\chi^2/2r},
\end{equation}
with the integrating factor $I(\chi) = \frac{1}{\chi}e^{-\chi^2/2r}$, where $c$ is a real constant that determines the shape of the orbit. Here we take the simpler case with $c=0$, to deal with analytical solutions. The orbit is then given by 
\begin{equation}
\label{Modified BNRT Orbit}
    \phi^2 + r\chi^2= 1 - 2r^2,
\end{equation}
which clearly requires that $r\in(0,\sqrt{2}/2)$. For simplicity, let us take $R=\sqrt{1-2 r^2}$.

\begin{figure}[!ht]
    \centering
\includegraphics[scale=0.3]{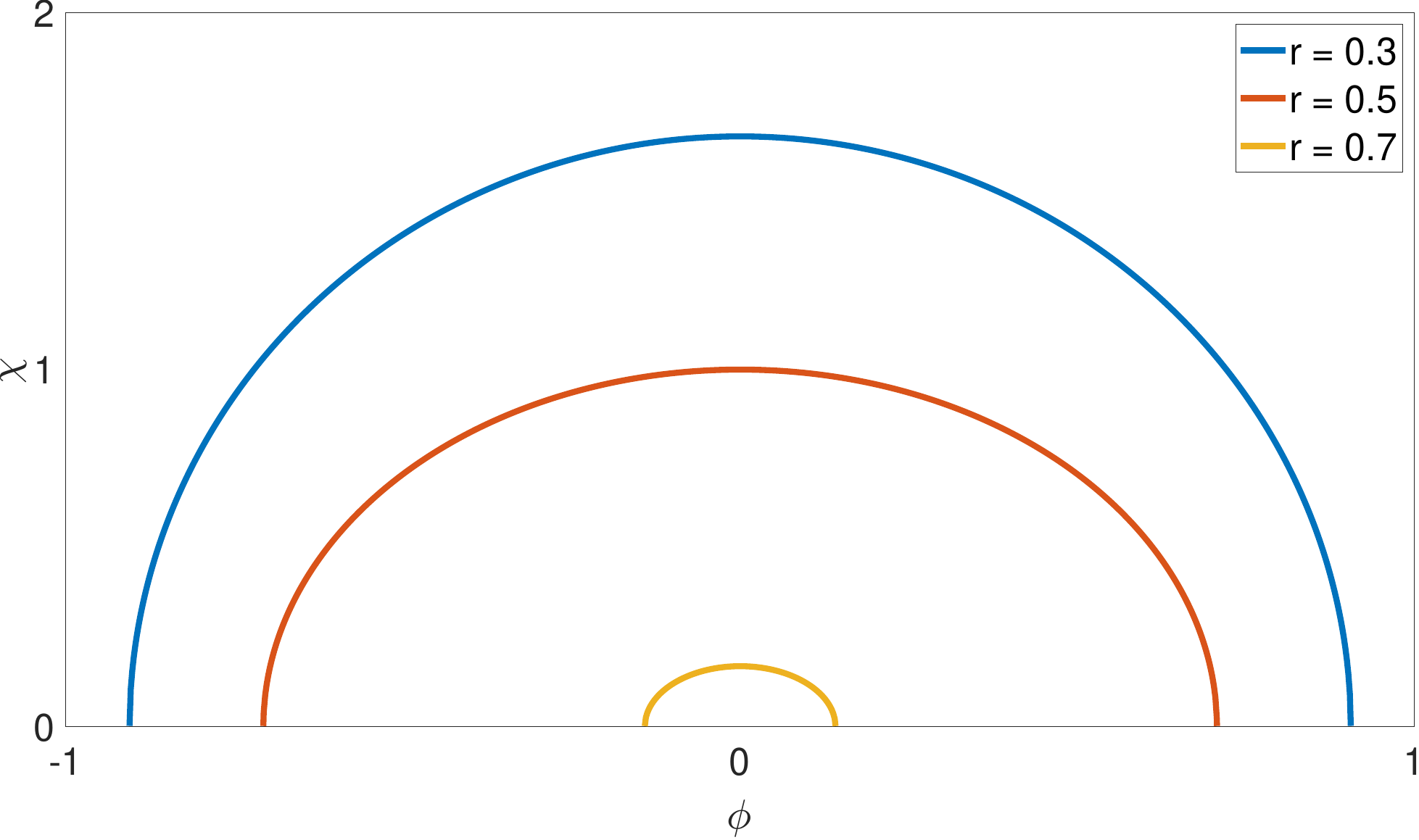}
    \caption{Some orbits for the modified BNRT model}
    \label{fig:Orbit for the Modified BNRT Model}
\end{figure}

\begin{figure}[!ht]
    \centering
    \includegraphics[scale=0.3]{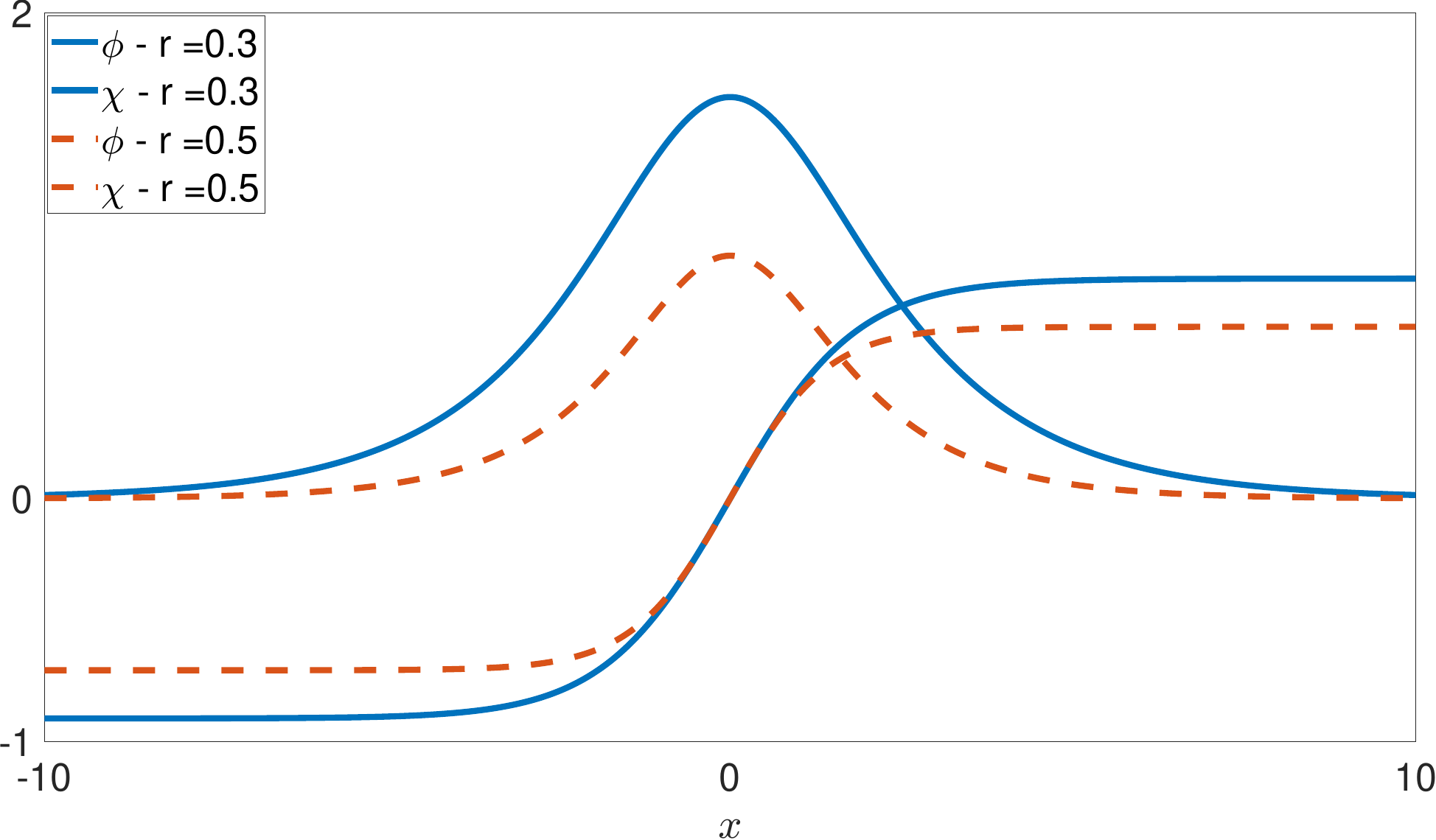}
    \caption{ The $\phi$ and $\chi$ solutions for the modified BNRT model, for $r=0.3$ (blue) and $0.5$ (orange)}
    \label{Solução - BNRT Modificado}
\end{figure}

In the present case, however, we can search for orbits that connect the pair of minima $(\pm R,0)$, which depend on the value of $r$. The profile of the orbit is shown in Fig. \ref{fig:Orbit for the Modified BNRT Model} for $r=0.3, 0.5$ and $0.7$. Substituting \eqref{Modified BNRT Orbit} in the first-order equation for $\phi$, we get
\begin{equation}
    \phi ' = 2r\PC{R^2-\phi^2},
\end{equation}
which has as solution
\begin{equation}
    \phi = R\, \tanh(2r R x),
\end{equation}
and the field $\chi$ is then given by
\begin{equation}
    \chi= \frac{R}{\sqrt{r}}\,\sech(2rR x).
\end{equation}
The solution profile is shown in Fig. \ref{Solução - BNRT Modificado}. We observe that the solution of the field $\phi$ doesn't connect asymptotically the minima $(\pm 1, 0)$ anymore, as it did in the standard BNRT model. Moreover, both $\phi$ and $\chi$ have the same width $\ell \propto  1/2rR $, and amplitude controlled by $r$, but they do not have the same amplitude for $r\in(0,\sqrt{2}/2)$, required by the above $\phi$ and $\chi$ solutions.

\begin{figure}[!ht]
    \centering
\includegraphics[scale=0.3]{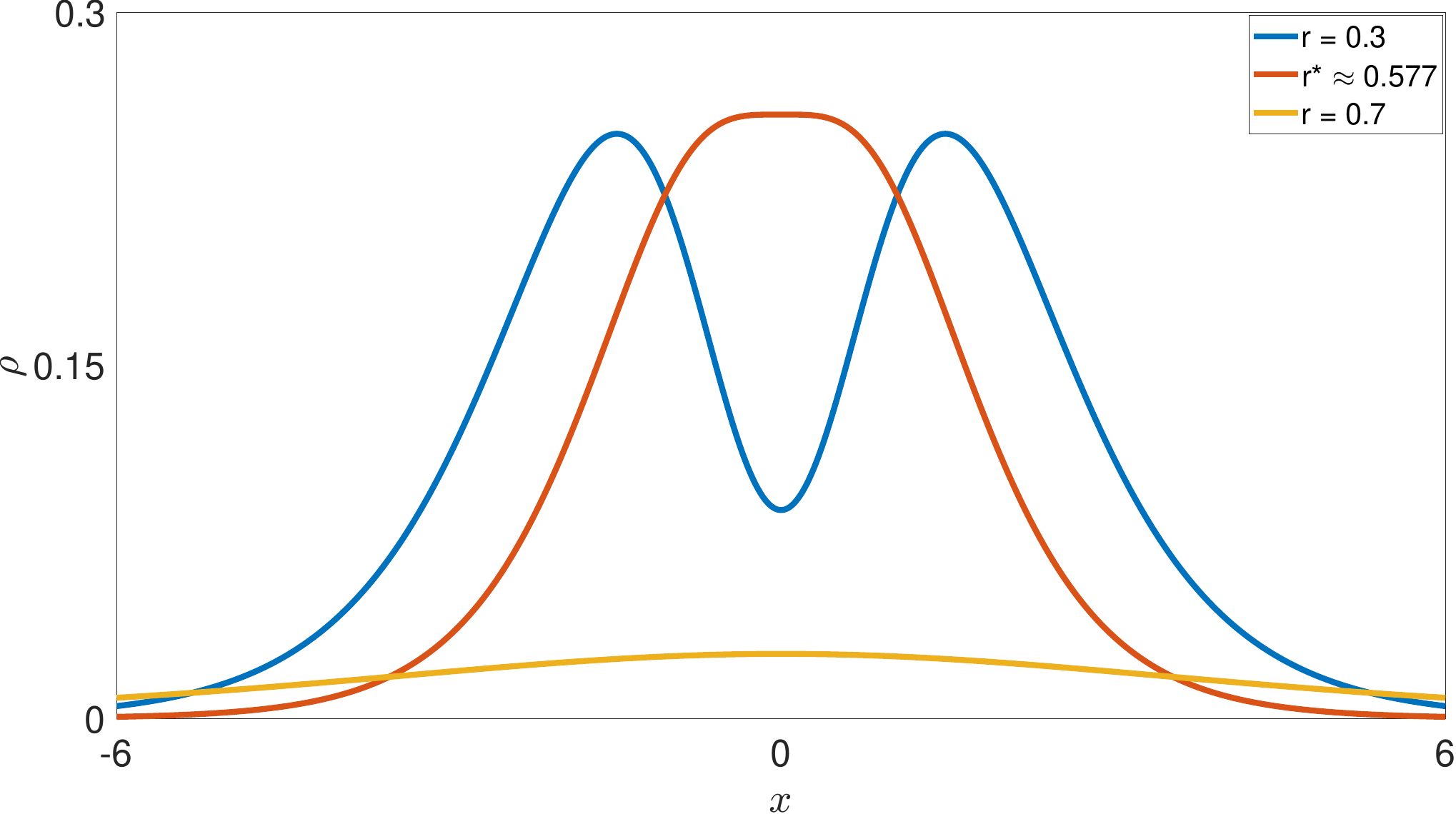}
    \caption{Energy density for the modified BNRT model}
    \label{fig:Energy Density for the Modified BNRT Model}
\end{figure}

\begin{figure}[!ht]
    \centering  \includegraphics[scale=0.3]{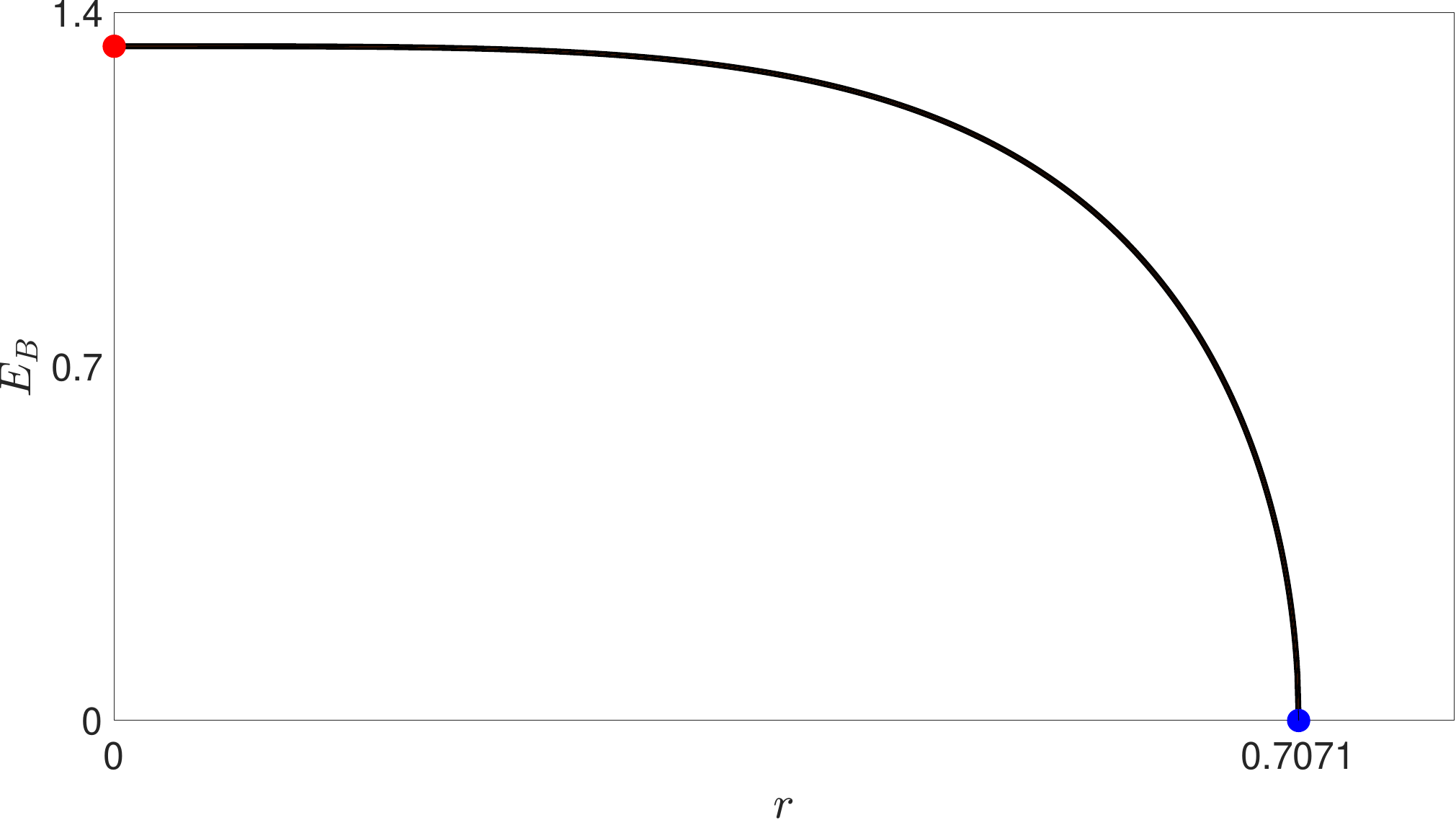}
    \caption{The energy for the solutions of the modified BNRT model.}
\label{fig:Energy for the Modified BNRT Model}
\end{figure}

The energy density is now given by
\be
    \rho = 4r R^2\sech^2(2rR x)[r^2 + R^2\tanh^2(2rR x)].
\ee
It is depicted in Fig. \ref{fig:Energy Density for the Modified BNRT Model} for some values of $r$. In particular, it also display the splitting phenomenon. 
Since we have established a first-order framework, the energy of this solution is given by $E_{B} = \vert W(\phi (\infty), \chi (\infty)) - W(\phi (-\infty), \chi (-\infty)) \vert$, which reads

\begin{equation}
    E_{B} = \frac{4}{3}R\,(1+r^2).
\end{equation}
The dependence of the energy on $r$ is shown in Fig. \ref{fig:Energy for the Modified BNRT Model}. There, the red and blue dots are depicted just to inform the corresponding asymptotic values, but they do not represent the modified BNRT model. Since the energy $E_{B} \neq 0$ in the interval in which $r$ is defined, we can ensure that the topological sector that connects the minima $(\pm R,0)$ is always BPS.

As we have already commented above, the energy density engenders an  interesting feature, the splitting at $r^{*}=0.577$, having two distinct maxima for $r\in(0,0.577)$. 
This behavior has also appeared in the BNRT model, but here $r$ includes a larger interval, with the amplitude of the $\phi$ field solutions now depending on $r$, so it is not fixed anymore. The two peaks or the bimodal form of the energy density inside the interval $(0,0.577)$ may be of interest in applications of practical use in other areas of nonlinear science, as we shall further comment below. 

The above model, which includes a geometric modification of the standard BNRT model, can be extended following the lines of \cite{Paga}, including another scalar field to control in internal structure of the localized solution. Moreover, it can be used within the context of spectral wall in multifield kink dynamics, as considered before in Ref. \cite{SW}, where a different modification of the BNRT model has been studied. Such possibility can also be directly added to the modified BNRT model just investigated, so we can study how the geometric modification that appeared above can compete with the distinct kinematical modification included in \cite{SW} to study spectral wall. This is of current interest, and may shed further light on the spectral wall phenomenon discovered recently in \cite{IM}.

\section{Ending Comments}\label{E}

In this work, we have investigated several models described by two real scalar fields $\phi$ and $\chi$ in $1+1$ spacetime dimensions. We have focused mainly on the case of modified dynamics, with the inclusion of the function $f(\chi)$ in the kinetic term of the field $\phi$. We started the present study with the methodology developed in Sec. \ref{M}, briefly reviewing the BNRT model and the model with modified dynamics, searching for first-order differential equations that solve the equations of motion.

 We have investigated several new systems in Sec. \ref{R}, in particular, the 4-6, the 6-4, the 6-6, and two others, introducing the vacuumless model as the V-4 and the V-6 models. Moreover, in Sec. \ref{IV} we have included the modified BNRT model, which brought an interesting new result, with analytical solutions even when the auxiliary function $W(\phi,\chi)$ is not separable as the sum of $W_1(\phi)+W_2(\chi)$. This have led  us to the much harder case in which the first-order equations are not separable anymore. However, we have used the method of the integrating factor to solve the problem when $f(\chi)=1/\chi^2$. The integrating factor opened the possibility to find analytical solutions, with the energy density displaying the interesting feature of adding the splitting behavior. Since the BNRT model leads to the presence of Bloch wall, the modified BNRT model may suggest the presence of modified Bloch wall, which can certainly be of practical use in applications to control the magnetization in magnetic materials, to name an interesting area of current interest. 

In this directions, in particular, we recall two distinct lines of investigations, one \cite{Paga} in which we can add distinct possibilities of modification of the internal structure of Bloch walls, and the other \cite{sky1,sky2,Rod2} in which we can simulate and control the magnetization inside skyrmions in magnetic elements.
The subject is related, for instance, to the very recent work \cite{AAA}, where the authors observed the formation of magnetic domain walls down to the single-layer limit
in the magnetically ordered van der Waals material CrSBr. With the methodology developed in \cite{sky1,sky2}, one can perhaps map planar skyrmion-like configurations and, with the geometric modification described in this work, control the magnetization inside the localized structure. As one knows, like domain walls, skyrmions can also be of the Néel or Bloch type (see, e.g., Refs. \cite{NNano,Skyr} and references therein), and we can further explore the skyrmion profile under the geometric constriction studied in the present work.

As we have seen from the systems studied in Sec. \ref{R}, in models where $W(\phi,\chi)=W_1(\phi)+W_2(\chi)$, the behavior of the field $\phi$ can be used to describe localized structures of the Néel type, since it does not contribute to change the behavior of the field $\chi$. However, when $W(\phi,\chi)$ is not separable, as described in Sec. \ref{IV}, the localized structure seems to engender the Bloch type profile, once the  fields $\phi$ and $\chi$ act collectively. This is an important identification, to be further explored elsewhere in the study of Bloch walls and skyrmions in magnetic materials. In application concerning the condensed matter framework, we can think of simulating the DM interaction adding another contribution to the kinematics of the two fields, in a way similar to the case already investigated in \cite{LV}, in the study of kinks in the presence of Lorentz violation. There, the authors implemented a nontrivial Lorentz breaking scenario by introducing an external, constant vector $k^\mu$, coupled with the two real scalar fields $\phi$ and $\chi$ in the form $\phi k^\mu\partial_\mu\chi$. The constant vector 
$k^\mu$ introduces a peculiar direction in the spacetime which breaks Lorentz invariance and induces an asymmetry in the field configurations that solve the equations of motion. This asymmetry is similar to the chiral asymmetry that appears under the DM interaction, thus motivating new investigation in the subject. As we can see, for static configurations, the term $\phi k^\mu\partial_\mu\chi$ becomes proportional to $\phi(d\chi/dx)$, and this is similar to the term included in \cite{SW} to investigate the presence of spectral walls in multifield kink dynamics. In this sense, the presence of Lorentz violation, which is related to the spectral wall effect, may also be connected to the DM interaction, bringing to the scene a novel connection between spectral wall and chiral excitation in magnetic materials. Furthermore, the relativistic models studied in Sec. \ref{R} and in Sec. \ref{IV} can be naturally extended to include curvature effects, changing the Minkowski metric considered in this work to other more general curvature background possibilities, in this sense allowing further contact with Refs. \cite{PRL,Toma,PRB}, to study the motion of localized magnetic structures in magnetic materials.

We can also think of other applications in high energy physics. An interesting possibility is to consider other models for the fields $\phi$ and $\chi$. This may be investigated using the deformation procedure developed in \cite{BLM} to construct new models, as recently described in Ref. \cite{Gani}, where the authors studied the presence of kinks with power-law tails. The study of localized structures with power law tail is of current interest and may be considered to describe interactions among systems such as Ridberg atoms and cold atoms in cavities; see, e.g., Ref. \cite{Rid} and references therein. Another possibility is to suppose different changes in the kinematics of one of the two fields. For instance, one could use the Dirac-Born-Infeld type of modification, which introduces distinct nonlinearities to the kinetic part of the Lagrangian, adding a square root restricting the field evolution and including additional powers in derivatives of the scalar field, controlled by a single real parameter. This line of investigation will bring new effects and could follow Ref. \cite{AoP}, which considered a single scalar field, leaving room for the case of multifield models. An additional route is to use the above results to investigate the case of Bloch brane considered before in Ref. \cite{B}. Here we can take the modified BNRT model to construct modified Bloch brane, in a way similar to that described in the work \cite{MB}, in which one suggested the presence of brane in the context of  constrained geometry studied in the present work. 
We can also consider the modified BNRT model in the presence of fermions, to see how the modified configuration may change the presence of bound states inside the fermionic gap, in a way similar to the study developed before in Ref. \cite{AMo}. This is also of current interest, since it may lead to information on current-induced interactions to generate and control the magnetization dynamics in magnetic materials \cite{RMP}. 

We are now considering some of the above possibilities, in particular, the study of kink scattering, to solve numerically the pair of equations \eqref{LE} and to see how the geometric constriction described by $f(\chi)$ may contribute to change the collision output, in a way similar to the case recently discussed in \cite{GC}. We are also considering the addition of the term $\phi\, k^\mu\partial_\mu\chi$ to the Lagrangian density \eqref{modify}, to study the presence of kinklike configurations that break the chiral symmetry and are of practical use in applications in magnetic materials, in connection with the spectral wall phenomenon recently considered in Refs. \cite{IM,SW}. We hope to report on some of these issues in the near future.

{{\bf Acknowledgments:} This work was partially financed by Coordenação de Aperfeiçoamento de Pessoal de Nível Superior (CAPES), Grant 88887.899555/2023-00 (GSS), by Conselho Nacional de Desenvolvimento Científico e Tecnológico (CNPq), Grants 303469/2019-6 (DB) and 310994/2021-7 (RM), by Fundação de Apoio à Pesquisa do Estado de Goiás, Grant 202110267000415 (MAF), and by Paraiba State Research Foundation, Grants 0003/2019 (RM) and 0015/2019 (DB).}

\bigskip

{\bf Data Availability Statement:} This manuscript has no associated data.


\begin{thebibliography}{99}
\bibitem{KM}Y. S. Kivshar and B. A.  Malomed, Rev. Mod. Phys. 61, 763 (1989).
\bibitem{Vi}A. Vilenkin and E. P. S. Shellard, {\it{Cosmic Strings and Other Topological Defects}} (Cambridge University Press, 2000).
\bibitem{Ma}N. S. Manton and P. Sutcliffe, {\it{Topological solitons}} (Cambridge University Press, 2004).
\bibitem{Va}T. Vachaspati, {\it{Kinks and Domain Walls: An Introduction to Classical and Quantum Solitons}} (Oxford University Press,
2007).
\bibitem{Sc}Y. M. Shnir, {\it{Topological and Non-Topological Solitons in Scalar Field Theories}} (Cambridge University Press, 2018).
\bibitem{NO}H. Nielsen and P. Olesen, Nuclear Physics B 61, 45 (1973).
\bibitem{F}J.-Z. Wu, H.-Y. Ma, and M.-D. Zhou, {\it{Vorticity and Vortex Dynamics}} (Springer, 2006).
\bibitem{Su}R. Huebener, N. Schopohl, and G. Volovik, eds., {\it{Vortices in Unconventional Superconductors and Superfluids}} (Springer,
2002).
\bibitem{BEA}C. J. Pethick and H. Smith, {\it{Bose-Einstein Condensation in Dilute Gases}} (Cambridge University Press, 2002).
\bibitem{BEB}V. S. Bagnato, D. J. Frantzeskakis, P. G. Kevrekidis, B. A. Malomed, and D. Mihalache, Romanian Reports Physics 67,
5 (2015).
\bibitem{tH}G. ’t Hooft, Nuclear Physics B 79, 276 (1974).
\bibitem{Po}A. M. Polyakov, JETP Lett. 20, 194 (1974).
\bibitem{Sch}Y. M. Shnir, {\it{Magnetic Monopoles}} (Springer, 2005).
\bibitem{Mil}K. A. Milton, Reports on Progress in Physics 69, 1637 (2006).
\bibitem{SI}C. Castelnovo, R. Moessner, and S. L. Sondhi, Nature 451, 42 (2008).
\bibitem{SI2}S. T. Bramwell, S. R. Giblin, S. Calder, R. Aldus, D. Prabhakaran, and T. Fennell, Nature 461, 956 (2009).
\bibitem{Malo}B. A. Malomed, {\it{The sine-Gordon Model: General Background, Physical Motivations, Inverse Scattering, and Solitons. In: Cuevas-Maraver, J., Kevrekidis, P. and Williams, F. (eds) The sine-Gordon Model and its Applications.}} (Springer, 2014).
\bibitem{Camp}D. K. Campbell, J. F. Schonfeld, and C. A. Wingate, Physica D 9, 1 (1983).
\bibitem{Const}D. Bazeia, M. A. Liao, and M. A. Marques, European Physical Journal Plus 135, 383 (2020).
\bibitem{sala}A. J. Balseyro Sebastian, D. Bazeia, and M. A. Marques, EPL 141, 34003 (2023).
\bibitem{tail}D. Bazeia, M. A. Marques, and R. Menezes, European Physical Journal Plus 138, 735 (2023).
\bibitem{prbrc}P. O. Jubert, R. Allenspach, and A. Bischof, Phys. Rev. B 69, 220410 (2004).
\bibitem{DHN}R. F. Dashen, B. Hasslacher, and A. Neveu, Phys. Rev. D 10, 4130 (1974).
\bibitem{Montonen}C. Montonen, Nuclear Physics B 112, 349 (1976).
\bibitem{STB}S. Sarker, S. E. Trullinger, and A. R. Bishop, Physics Letters A 59, 255 (1976).
\bibitem{Raj}R. Rajaraman, Phys. Rev. Lett. 42 (1979).
\bibitem{Ruck}H. M. Ruck, Nuclear Physics B 167, 320 (1980).
\bibitem{JR}R. Jackiw and C. Rebbi, Phys. Rev. D 13, 3398 (1976).
\bibitem{SSH}W. P. Su, J. R. Schrieffer, and A. J. Heeger, Phys. Rev. Lett. 42, 1698 (1979).
\bibitem{Hee}A. J. Heeger, Rev. Mod. Phys. 73, 681 (2001).
\bibitem{Mansfield}M. L. Mansfield, Chemical Physics Letters 69, 383 (1980).
\bibitem{Ventura}D. Bazeia and E. Ventura, Chemical Physics Letters 303, 341 (1999).
\bibitem{GW}J. Goldstone and F. Wilczek, Phys. Rev. Lett. 47, 986 (1981).
\bibitem{AMo}D. Bazeia, A. Mohammadi, and D. C. Moreira., Phys. Rev. D 103, 025003 (2021).
\bibitem{All}A. Vanhaverbeke, A. Bischof, and R. Allenspach, Phys. Rev. Lett. 101, 107202 (2008).
\bibitem{Net1}P. Fendley, S. D. Mathur, C. Vafa, and N. P. Warner, Physics Letters B 243, 257 (1990).
\bibitem{Net2}E. R. C. Abraham and P. K. Townsend, Nuclear Physics B 351, 313 (1991).
\bibitem{Net3}V. I. Afonso, D. Bazeia, M. A. G. León, L. Losano, and J. M. Guilarte, Nuclear Physics B 810, 427 (2009).
\bibitem{G}G. W. Gibbons and P. K. Townsend, Phys. Rev. Lett. 83, 1727 (1999).
\bibitem{Sa}P. M. Saffin, Phys. Rev. Lett. 83, 4249 (1999).
\bibitem{BB}D. Bazeia and F. A. Brito, Phys. Rev. Lett. 84, 1094 (2000).
\bibitem{P6}P. Dorey, K. Mersh, T. Romanczukiewicz, and Y. Shnir, Phys. Rev. Lett. 107, 091602 (2011).
\bibitem{LR}I. C. Christov, R. J. Decker, A. Demirkaya, V. A. Gani, P. G. Kevrekidis, A. Khare, and A. Saxena, Phys. Rev. Lett.
122, 171601 (2019).
\bibitem{IM}C. Adam, K. Oles, T. Romanczukiewicz, and A. Wereszczynski, Phys. Rev. Lett. 122, 241601 (2019).
\bibitem{SW}C. Adam, K. Oles, T. Romanczukiewicz, A. Wereszczynski, and W. J. Zakrzewski, Journal of High Energy Physics 08,
147 (2021).
\bibitem{GC}J. G. F. Campos, F. C. Simas, and D. Bazeia, Journal of High Energy Physics 10, 124 (2023).
\bibitem{MD}A. Hubert and R. Sch{\"a}fer, {\it{Magnetic Domains: The Analysis of Magnetic Microstructures}} (Springer, 1998).
\bibitem{sky}A. Fert, V. Cros, and J. Sampaio, Nature Nanothechnology 8, 152 (2013).
\bibitem{sky1}D. Bazeia, M. M. Doria, and E. I. B. Rodrigues, Physics Letters A 380, 1947 (2016).
\bibitem{sky2}D. Bazeia, J. G. G. S. Ramos, and E. I. B. Rodrigues, Journal of Magnetism and Magnetic Materials 423, 411 (2017).
\bibitem{D}I. E. Dzyaloshinskii, J. Phys. Chem. Solid. 4, 241 (1958).
\bibitem{M}T. Moriya, Phys. Rev. 120, 91 (1960).
\bibitem{DMInew}R. E. Camley and K. L. Livesey, Surface Science Reports 78, 100605 (2023).
\bibitem{PRL}Y. Gaididei, V. P. Kravchuk, and D. D. Sheka, Phys. Rev. Lett. 112, 257203 (2014).
\bibitem{Toma}R. Tomasello, E. Martinez, R. Zivieri, L. Torres, M. Carpentieri, and G. Finocchio, Scientific Reports 4, 6784 (2014).
\bibitem{PRB}K. V. Yershov, V. P. Kravchuk, D. D. Sheka, O. V. Pylypovskyi, D. Makarov, and Y. Gaididei, Phys. Rev. B 98, 060409(R)
(2018).
\bibitem{Bogomol'nyi}E. B. Bogomolny, Sov. J. Nucl. Phys. 24, 449 (1976).
\bibitem{BSR}D. Bazeia, M. J. dos Santos, and R. F. Ribeiro, Physics Letters A 208, 84 (1995).
\bibitem{BNRT}D. Bazeia, J. R. S. Nascimento, R. F. Ribeiro, and D. Toledo, Journal of Physics A 30, 8157 (1997).
\bibitem{IntegratingFactor}A. A. Izquierdo, M. A. G. Le\'on, and J. M. Guilarte, Phys. Rev. D 65, 085012 (2002).
\bibitem{Vil}I. Cho and A. Vilenkin, Phys. Rev. D 59, 021701 (1998).
\bibitem{Vacuumless}D. Bazeia, Phys. Rev. D 60, 067705 (1999).
\bibitem{Paga}D. Bazeia, M. A. Marques, and M. Paganelly, European Physical Journal Plus 137 (2022).
\bibitem{Rod2}D. Bazeia, D. C. Moreira, and E. I. B. Rodrigues, Journal of Magnetism and Magnetic Materials 475, 734 (2019).
\bibitem{AAA}Y. Zur, A. Noah, C. Boix-Constant, and et al., Adv. Mater. 35, 2307195 (2023).
\bibitem{NNano}N. Nagaosa and Y. Tokura, Nature Nanotech. 8, 899 (2013).
\bibitem{Skyr}W. Kang, Y. Huang, X. Zhang, I. Zou, and W. Zhao, Proc. IEEE 102, 2040 (2016).
\bibitem{LV}M. N. Barreto, D. Bazeia, and R. Menezes, Phys. Rev. D 73, 065015 (2006).
\bibitem{BLM}D. Bazeia, L. Losano, and J. M. C. Malbouisson, Phys. Rev. D 66, 101701(R) (2002).
\bibitem{Gani}P. A. Blinov, T. V. Gani, and V. A. Gani, Annals of Physics 437, 168739 (2022).
\bibitem{Rid}N. Defenu, T. Donner, T. Macrì, G. Pagano, S. Ruffo, and A. Trombettoni, Rev. Mod. Phys. 95, 035002 (2023).
\bibitem{AoP}D. Bazeia, E. E. M. Lima, and L. Losano, Annals of Physics 388, 408 (2018).
\bibitem{B}D. Bazeia and A. R. Gomes, Journal High Energy Physics 05, 012 (2004).
\bibitem{MB}D. Bazeia, D. A. Ferreira, and M. A. Marques, European Physical Journal Plus 135, 587 (2020).
\bibitem{RMP}A. Manchon, J. Železný, I. M. Miron, T. Jungwirth, J. Sinova, A. Thiaville, K. Garello, and P. Gambardella, Rev. Mod. Phys. 91, 035004 (2019).
\end{thebibliography}
\end{document}